# *Geoastronomy*: Rocky planets as the *Lavosier-Lomonosov Bridge* from the non-living to the living world

Stephen J. Mojzsis[a,b,c*]

[a] Origins Research Institute, Research Centre for Astronomy and Earth Sciences, H-1112 Budapest, Hungary

[b] Department of Lithospheric Studies, University of Vienna, 1090 Vienna, Austria

[c] Department of Geological Sciences, University of Colorado, Boulder, CO 80309-0399, USA

*corresponding email address: smojzsis@gmail.com

*DEDICATION:*

*To Gustaf O. Arrhenius, who showed me that chemistry need not be solely bound to Earthly topics.*

## ABSTRACT/WEB SUMMARY

Life on Earth emerged at the interface of the planet's geosphere, hydrosphere and atmosphere. This setting serves as our basis for how biological systems originate on rocky planets. Often overlooked, however, is the fact that a terrestrial-type planet's chemical nature is ultimately a product of the Galaxy's long term evolution. Elemental abundances of the major rock-forming elements (e.g. Si, Mg, Fe) can be different for different stars and planets formed at different times in galactic history. These differences mean that we cannot expect small rocky exoplanets to be just like Earth. Furthermore, age of the system dictates starting nuclide inventory from galactic chemical evolution, and past, present and future mantle and crust thermal regimes. A rocky planet's bulk silicate mantle composition modulates the kind of atmosphere and hydrosphere it possesses. Hence, the *ingredients* of a rocky planet are as important for its potential to host life as proximity to the so-called *habitable zone* around a star where liquid water is stable at the surface. To make sense of these variables, a new trans-disciplinary approach is warranted that fuses the disciplines of Geology and Astronomy into what is here termed, *Geoastronomy*. **(190 words)**

## 1. Introduction.

What makes a relatively small silicate+metal (i.e. "rocky") planet like Earth, orbiting a Main Sequence dwarf star like the Sun, suitable for the origin and propagation of life? A criterion commonly used to put boundary conditions on the answer to this question is the so-called "habitable zone" (HZ) concept whereby liquid water may remain stable over billion year (Gyr) timescales on planetary surfaces under certain conditions. The habitable zone has as its key input parameters: Mass of the planet; distance from the star; albedo; atmospheric greenhouse inventory that regulates surface warming from insolation; and, abundant available liquid water as an initial condition (1-4). In the Solar System, Earth is at an optimal location for fulfilling these criteria over geologic timescales, whereas Mars has always resided at the outer edge of this zone, and Venus is situated at its Sun-ward inner edge. What has been dubbed the "Goldilocks planet" concept is straightforward, and has a surprisingly long history (e.g. I. Newton in 1687[1]), before beginning its modern development in the early 1900s (5). Yet, despite the obvious historical importance of the HZ in influencing how we search for life in the universe, the modern application of the concept is weakened when confronted with the appropriate astro- and geo-physical caveats, foremost of these being that the habitable zone has greatly changed with time as the Sun's luminosity has evolved (6). Furthermore, an impulse that follows from placing much faith in this simple idea, is that in order for us to find life in the universe we ought to "Follow the Water" (7). The HZ approach is also undermined by the fact that the Solar System's liquid water habitable volume is not actually defined by the familiar naked seas and transparent skies of Earth. Rather, most of the potential habitable volume in our Solar System is to be found in the cloaked oceans locked within the largest Trans-Neptunian Objects at the edge of interplanetary space (8). These worlds are akin to the postulated cold and sunless bodies that wander between the stars (9) (**Figure 1**). Furthermore, some exoplanets in their habitable zones around Main Sequence stars could have inherited very little or no water at all, whereas others may have ocean inventories factors of tens or hundreds more than Earth's. We are left with the humbling prospect that searches for life on an Earth 2.0 somewhere else actually leave the most numerous habitats of all in the dark.

[1] Isaac Newton: *Philosophiae Naturalis Principia Mathematica*. 3rd Ed., Book III, Section I, p. 722. Translated and Annotated by Ian Bruce (1728).

<INSERT FIGURE 1 HERE>

That being said, water is common in the cosmos. It is composed of two ($_1$H, $_8$O) of the three most abundant nuclides in the universe (the other being $_2$He). It is stable in liquid form under definite Pressure and Temperature conditions present in a surprisingly wide variety of planetary environments on Earth, in and on Mars, and even in Venus' atmosphere (**Figure 2**). Concerns over the availability of liquid water that compel us to "Follow the Water" in the search for life beyond Earth turn out to be less important than achieving a better understanding of the bulk physical chemistry of the planetary object itself that may or may not host water. With this understanding, our focus abruptly shifts to quantitative understanding how a terrestrial planet, icy moon, or even rogue planet untethered to its star, comes to have the characteristics it does rather than embarking on arguments over what habitability actually means.

<INSERT FIGURE 2 HERE>

The logical basis for sustaining life on small rocky planets that instead "Follows the Chemistry" – whether or not such life is actually there or not – is a notion advocated by the 18th century CE physical chemists A. Lavoisier and M.V. Lomonosov (10). Both remarked on the bewildering complexity of living matter – what was termed by them at that time as the "animate world" – which is nevertheless based on a small number of elements. They contrasted life chemistry with the relatively straightforward physical chemistry of solids, especially minerals, that comprise what was termed the "inanimate world". Minerals are natural crystalline solids of definite composition and structure that spontaneously form from all of the naturally occurring elements (surprisingly, this includes some of the noble gases; e.g. ref. 11). Furthermore, the first minerals in the universe crystallized from the earliest nucleosynthetic products of common baryonic matter[2]: Of these, Carbon gave rise to graphite/diamond, and 2H+O ($H_2O$) to water ice (12). The chemical difference between the animate/inanimate world therefore hints that rocks and minerals – naturally occurring solid substances with repeating charge structure – represent the linkage between non-life and life. That linkage is here termed the *Lavosier-Lomonosov Bridge*.

### 1.1 Follow the (cosmo)chemistry

The *Lavosier-Lomonosov Bridge* hypothesis posits that the planets and the rocks and minerals that comprise them nurtured chemistry of a small number of common nuclides: $_6$C, $_1$H, $_8$O, $_7$N, $_{16}$S and $_{15}$P that led to the complexity of biology. These six elements are also widely present in different minerals in planetary crusts, in hydrospheric fluids and some are also found in atmospheric gases. Overall, of the 90 stable and long-lived (>$10^9$ y) nuclides, only about 21 are directly essential to biochemistry. Those that play the most important chemical roles in biology are also of relatively low atomic number; only two ($_{42}$Mo, $_{53}$I) are above that of $_{34}$Se (**Figure 3**). The four most abundant elements in living organisms, in terms of percentage of total number of atoms, are: hydrogen, oxygen, nitrogen, and carbon, which together make up more than 99% of the mass of a living cell. Besides their universal abundance, the top four biophilic elements are also the lowest atomic number nuclides capable of forming 1-, 2-, 3-, and 4-bonds, respectively. The significance of these observations for understanding life in the universe cannot be overstated. Along these lines, some have argued that owing to the collective traits cited above, life's chemistry can be confidently viewed as universal (13). The view taken in the work presented here is more agnostic, but nevertheless it is appropriate that we should begin our search for life elsewhere from the chemistry that we know while acknowledging that we will be in for surprises when life elsewhere is discovered.

[2] We do not know what Dark Matter is (~27% of the mass of the universe), much less whether it could have intrinsic structure like minerals do.

<INSERT FIGURE 3 HERE>

Hydrogen was formed in Big Bang nucleosynthesis, whereas oxygen, nitrogen and carbon are commonly produced in nucleosynthesis by dying low-mass stars and exploding massive stars. Sulfur and Phosphorus on the other hand are synthesized in exploding massive stars and some fraction, by exploding white dwarfs (14). In light of this history, we can conclude that the universe ca. 13.7 Ga did not begin with the elements deemed necessary for the chemistry of life as we know it, but had to nucleosynthetically evolve through multiple generations of stars in galactic environments such as our own to provide the current biophilic inventory (**Figure 4**). It is currently unknown when this inventory was reached in galactic history to have first facilitated the *Lavosier-Lomonosov Bridge*.

<INSERT FIGURE 4 HERE>

Following the chemistry back to planets, a rocky world's elemental makeup governs nearly all of its major processes, just as the chemistry of life defines the processes of living organisms. Planets are gigantic chemical masses, and the bulk composition of a terrestrial planet modulates the nature of its mantle, and therefore its crust, atmosphere and hydrosphere. For example, without Earth's globally active crustal-recycling process, plate tectonics, dictated by internal heat, water content and mineral composition (expressed as rheology[3]), a sustained carbon cycle and thus long-term habitability of our planet is generally deemed unlikely (e.g. 15). To better understand worlds like Earth around other stars (exoplanets) that could harbour the conditions necessary for prebiotic chemistry to lead to life, we must understand how differences in planet compositions arise and lead to different evolutionary tracks.

The problem of course is that we do not know enough about rocky exoplanets and their differences to say much about possible divergent evolutionary tracks that could exist. Owing to this ignorance, discoveries of rocky exoplanets have, up to now, inspired diverse ideas about plausible structures and tectonic regimes that bear on their proclivity for life. Such studies are severely hampered by inexact assumptions about intrinsic radiogenic heat production and cosmochemically plausible inventories of rock-forming elements that went into forming them. Galactic Chemical Evolution (GCE) models provide some information to overcome this limitation. Such GCE models track star (and planet) age and composition trajectories that describe different physically possible compositions that in turn effect geodynamical regimes. These further dictate whether or not a world has the potential to originate the kind of life as we know it and to sustain this life over billions of years. We now ask ourselves: What are the boundary conditions for the origin of biochemistry on a planet in a cosmochemical backdrop?

[3] Rheology in the context of the interior of planets, is the study of the plastic (rather than elastic) flow of deformable solids such as mantle minerals under high pressure and temperature.

## 1.2 Guideposts for life

A leading premise in origins of life research is that biological systems will arise given the availability of 3 key environmental conditions on (or in) a planetary body (16).

These are:

(i) Abundant organic chemical products to build a (proto)biochemical scaffolding,
(ii) Reliable free energy from chemical disequilibria, thanks to electron flow from a reducing interior (mantle) to an oxidizing exterior (atmosphere),
(iii) Stable liquid water as the medium for the nascent biochemistry

and we now add to this list another important criterion,

(iv) Adequate time for Chemical Evolution to proceed to a degree of complexity that makes Darwinian Evolution possible.

Two important caveats to this necessarily brief (and admittedly, derivative) list are: First, that the initial steps to life are the most difficult (17); and, second, we do not know how long it takes to cross the *Lavosier-Lomonosov Bridge* even if all other conditions are met, ref. 18). Remarkably little attention has been paid to the temporal component (point iv on the above list), but it seems inarguable that the chemistry leading to life requires time to reach an adequate level of informational complexity to the point where a chemical entity is capable of being a biological entity that experiences Darwinian evolution (19). Taken together, a planet must have the capacity to maintain conditions (i-iii) long enough (iv) to allow Chemical evolution to achieve a certain intrinsic informational complexity that can be passed to the next generational chemical reaction in perpetuity. This last stage fulfils the ancestor-descendant relationship and thus represents the most simple definition of the transition from the non-living to the living.

It is unknown whether or how the appropriate informational complexity be reached in a chemical system without an underlying support mechanism (see, for example, ref. 20). To emphasize this point, T. Dobzhansky (1973; ref. 21) famously pointed out that "(n)othing in biology makes sense except in the light of evolution." S.L. Miller espoused the related view that the origin of life is the origin of (Darwinian) evolution (22). Whereas for our purposes, a plausible corollary to Miller's statement is that "the origin of evolution is the origin of information". We could say that the crustal platforms of planets, rocks and minerals, bathed in liquid water and atmospheric gases, provide the ultimate informational template if minerals were indeed the inorganic channel from mere chemistry to biochemistry. To understand the propensity for a biosphere to exist beyond Earth, we need to understand planets. To understand planets, we need to understand rocks. Taking a note from Dobzhensky, it is evident that *nothing in geology makes sense except in light of the physical and chemical properties of natural solids*. Thus, the science of life's origins should not be treated in a parochial fashion. As such, thoughts on the Origin of Life have never been exclusive to biology, or for that matter to chemistry, or even metaphysics/religion.

Notions about the emergence of the living world has had a strong foundation in geology, even from antiquity. Dramatic early examples are that of the Ionian Greek philosophers Θαλῆς (id. Thales) and Ἀναξίμανδρος (id. Anaximader), both of Miletus (ca. 5th c. BCE) who may have been the first to express in written form the concept that Earth and Life formed a *synergoi* [4], that an inexorable synergism exists between the origin of Earth and of the life on it. In the poem *On Nature* composed by Censorinus (3rd c., CE), Anaximander is said to have postulated that life's spontaneous generation befell "…in the residue of mud and mist on the Earth while the first water was being evaporated by the heat of the Sun" [5]. This remarkable

---

[4] https://en.wikipedia.org/wiki/Synergism_%28theology%29

[5] from: *De Die Natale, by Censorinus*, translated into English by William Maude, New York: The Cambridge Encyclopedia Co., 1900; pp. 9 to 26

statement presages by ca. 2400 years the oft-quoted correspondence of C. Darwin to J.D. Hooker (1871) that likewise invoked a “warm little pond” heated by the Sun [6] for the birthplace of life. By the end of the 19th c. CE, work had started to appear that assessed the search for chemical evidence of life in the oldest rocks (23). Later, *bona fide* experimental explorations of life’s origin commenced with the chemist W. Löb in 1906 (24,25) and reached its first spectacular milestone with the results of Miller in 1953 (26) who in a static discharge apparatus sought to recreate the conditions for prebiotic chemistry using what was considered at that time to be plausible geological conditions for the “Early Earth” that are different from that of today (27).

A starting point is the notion that conditions at the surface and interior of the Hadean Earth (before 4 Ga) which hosted life’s origin were different from those at present, but for reasons that are as yet poorly understood. It is generally accepted that the Hadean Sun was about 30% dimmer, its mantle hotter, the bombardment by asteroids and comets were frequent, and atmosphere anoxic (28). Little else is known, however, about how such differences influenced the way the young Earth functioned. By some accounts, the late Hadean (ca. 4.25-4.0 Gyr ago) may have witnessed the emergence of the biosphere from the ruins of planetary-scale bombardment (29,30). Other workers assert that life came much later. With regards to actually defining what early terrestrial environments existed that could have led to biological systems, and what environments did not, a huge number of fundamental aspects of the primordial Earth (e.g. temperature, continental:oceanic crust ratio, ocean volume, atmospheric composition and density) remain under debate. The modern roots of this debate stem in part from three deeply ingrained scientific concepts which emerged from Miller/Urey, and continue to strongly influence origins of life research (22; cf.ref. 31).

These enduring ideas can be summarized as follows:

(i) That a spontaneous origin of life occurred in natural geological environments intrinsically conducive to the generation of prebiotic organic molecules such as amino acids and nucleotides (e.g. adenine; ref. 32). Classical chemical thought posits that these molecules formed in a hydrogen-rich (reducing, electron-rich) setting with available cyanide, ammonia and other compounds that produced a “Prebiotic Soup” of simple molecules;

(ii) The existence of certain (geo)chemical, atmospheric and thermal conditions which concentrated organic molecules to reactivity, or facilitated reactions that gradually increased in complexity; and,

(iii) The pathways from organic chemical complexity via assembly of molecules to biology were episodically stymied because Earth’s surface took very long to cool after planet formation owing to impact bombardments by asteroids and comets.

As often occurs with the scientific enterprise, the concepts listed above to justify prebiotic experimental conditions were challenged almost immediately after they were first suggested.

### 1.2.1 *Miller-Urey vs. geochemistry*

Geologists consider it improbable that the Earth’s early atmosphere was ever very abundant in hydrogen, methane, ammonia and other reduced species as, for example, championed by Urey (27), and based on the earlier theoretical work of Haldane in 1929 (33) and Oparin in 1936 (34). The idea that life owes its existence to an original Jupiter-like atmosphere that shrouded the young Earth was viewed as radical even by Earth scientists in the late 1940s and well into the 1960s (35-37). Urey’s idea for a captured solar composition (H+He) gas in the Earth’s *primary atmosphere* at time of accretion is undermined by the fact that this gas should be lost to space on timescales of days to centuries (38). The Earth’s *secondary atmosphere*, outgassed from the cooling mantle, replaced the primary atmosphere very early (~40 Myr after planet formation; 39). Unless Earth’s primordial mantle was radically different from today's, which would have required changes to its chemical composition, the secondary atmosphere that was degassed could never have been Jupiter-like.

[6] https://www.darwinproject.ac.uk/entry-7471

In fact, our planet's mantle ended up being relatively oxidized as a consequence of heterogeneous accretion of planetary building blocks, and the chemical and mechanical separation of silicate vs. metal at the pressure-temperature conditions that existed in its own differentiation (40,41). This viewpoint was verified thanks to a record for Earth's mantle redox, or oxygen fugacity, from analysis of ancient zircons up to about 4.3 Gyr old (42). Oxygen fugacity is widely used as a measure of the degree of oxidation of rocks and the minerals that comprise them. Like other chemists, geochemists define fugacity (f) as the effective partial pressure of a gas (in this case, oxygen gas, ($fO_2$) in thermodynamic equilibrium with a particular mineral assemblage. Geochemists report oxygen fugacity, however, as the (non-ideal) partial pressure of $O_2$ relative to a convenient reference value such as graphite (C), metals, metal oxides and silicates. For Earth and other silicate+metal planets (and the Moon), ($fO_2$) tends to be reported in $\log_{10}$ units relative to the mineral buffers Iron-Wüstite[7] (IW) or Quartz- Fayalite-Magnetite, QFM). The intrinsic $fO_2$ of a terrestrial-type planet's mantle is an important parameter because it controls the relative proportions of FeO (wüstite) and $Fe_2O_3$ (ferric-oxide, hematite) and the stability of metallic Fe ($Fe^0$). In turn, variations in intrinsic $fO_2$ of a rocky planet can lead to different core:mantle ratios, and different chemical makeup of its mantle and thus of its crust, which in turn modulates the composition of de-gassed C-O-H components into its secondary atmosphere. Under reducing conditions, the principal exhaled gases are $H_2$ and CO, which at cooling react to form $H_2O$, $CH_4$ and $H_2$ (**Figure 5**), as opposed to the oxidizing conditions that prevail in Earth's upper mantle where $H_2O$ and $CO_2$ are the major species. Earth's lower mantle (25-50 GPa) is expected to have log $fO_2$ (Δ IW) around -1.5 (Frost & McCammon, 2008) but the upper mantle has apparently been close to its present oxygen fugacity (buffered at Fayalite-Magnetite-Quartz) for the last 4.3 billion years so that gaseous species exhaled from volcanoes forming the secondary atmosphere are dominated by $CO_2$, $H_2O$, $SO_2$, and $N_2$ (43,44). This observation does not bode well for an *endogenous* H-rich (reducing) Miller/Urey-type atmosphere on Hadean Earth (cf. 45).

<INSERT FIGURE 5 HERE>

We still do not know, however, whether the Earth's mantle was homogeneous with respect to oxygen fugacity or witnessed induced and transient heterogeneity due to *exogenous* effects of late accretion bombardment by metallic asteroids, and/or if it experienced secular redox changes in the deep past that have yet to be documented in the geologic record. Some workers in Origins of Life problems back-argue the point (46) along the lines that "…since there was an origin of life, the early atmosphere had to have been reducing." On the contrary, if our understanding is correct, silicate-metal planets of somewhat larger than Venus' mass (0.81 $M_\oplus$) are generally expected to grow progressively oxidized through a combination of metal sequestration in burgeoning Fe-Ni cores, $Fe^{2+}$-disproportionation during core growth, and the late arrival of volatile-rich and more oxidized planetesimals. As shown in Figure 5, smaller rocky planets such as Mars (0.1 $M_\oplus$) should have more reduced mantles that degas photochemically reactive gases such as $CH_4$. All other aspects being held equal, relatively oxidized mantles are probably the norm for terrestrial planets close to Earth's mass and composition, even those around other stars.

If there never really was a Prebiotic Soup, then the early ideas of Haldane, Oparin and Urey that organic compounds from different sources were dissolved in the oceans reacted to make increasingly complex compounds that fed a Heterotrophic origin of life must be either seriously re-evaluated (47), or abandoned. A variably dilute version of the Prebiotic Soup does not help either, because dilute solutions favour the breaking apart of molecules, not formation. On planets like Earth that have vast ($1.4 \times 10^{21}$kg; $1.335 \times 10^9$ km$^3$) oceans, there needs to be some way to concentrate organic molecules. Production of prebiotic chemical compounds in the oceans always risk becoming diluted into oblivion (48).

[7] Wüstite is simple ferrous-oxide; $Fe^{2+}O$. Fayalite ($Fe_2SiO_4$) is the $Fe^{2+}$-rich end-member of the olivine solid-solution series; Magnetite is the mixed valence ferro-ferric oxide ($Fe_3O_4$).

### 1.2.2 *Darwin ponds*

Beginning in the early 2000s, a number of published works on Origin of Life began to make the claim that there was "no dry land" on Earth at the time of life's origin and therefore life's origin by concentrated reactants in emergent crustal environments such as lakes, ponds, or shallow aquifers, was a dead-end. Without such emergent land, other mechanisms have been proposed to facilitate concentration such as mineral adsorption in solutions with dilute components (49,50). The debate about emergent land can be reduced to the deliberation over whether (small? medium? large?) granitic masses of buoyant crust could have existed in Earth's Hadean eon (ca. 4.5-3.85 Ga). In the last decade, work from Hf and Nd isotopes on the oldest rocks has shown that the basaltic-granitic dichotomy that typifies our planet as "ocean" and "land" has been around since about 4.45 billion years ago (51,52).

Planets like Earth have to lose heat convectively brought by their mantles to thermal boundary layers (i.e. crusts). Furthermore, on Earth and in the presence of abundant surface water (53), this heat loss process tends to generate continental crust (54). Surprisingly, granitic-type rocks are not unique to Earth; granodiorites have been found on Mars (55) and some granophyre-like inclusions occur in meteorites and lunar rocks (56). The origins of these extraterrestrial "granites" is poorly known but it shows that they are not unique to Earth. Dry land exists on Earth because of the density contrast between granite (2640 kg $m^{-3}$) and basaltic oceanic crust (2900 kg $m^{-3}$) riding on denser mantle (peridotite and dunite density ~3300 kg $m^{-3}$; average density ~5500 g $m^{-3}$). Heat loss happens via conduction through the thermal boundary layer (generally ineffective owing to the low thermal conductivity of rock; $k$ = 0.40–7.00 W $^{m-1}$ $K^{-1}$ 57), by mantle plumes that punch through the crust and create volcanic edifices (e.g. Hawai'I on Earth, Olympus Mons on Mars), or by plate tectonics. Only on Earth is heat-loss from plate tectonics operative, and it begins at the ~60,000 km-long mid-ocean ridge spreading axes and ends (ultimately) at subduction zones.

Consider that although the volume of Earth's continental crust may have been much less in the Hadean (perhaps 5% present continental area, or equivalent to about 90% of the size of Australia) there are opportunities for emergent land that are not associated whatsoever with continents. For instance, if the early Earth had shallower mean ocean depths of 2 km in the absence of continents to displace ocean water, it would mean that overall the mid-ocean ridge axes and plume-related oceanic volcanoes ("tropical Icelands") ought to be subaerial. Many contemporary islands rise above the ocean surface over mantle plume heads impinging oceanic crust (e.g. Hawai'I; Samoa, Tahiti, Easter, Azores, Iceland, Canary, Ascencion, Galapagos, Cocos). All of these exist today with an average ocean depth of 3.9 km. If ocean depth was even greater in the Hadean (by a factor of 2, but there is little or no evidence for this) it does not change the result. It is worth stating that opportunities for concentration exist on emergent land even on relatively water-rich Earth-like planets, including Super-Earths (58). Water World exoplanets of greater than 2 Earth radii (59) may be too hostile to the chemistry of life. If endogenous conditions favour life on a terrestrial-type planet, what threats are posed to a nascent biosphere by asteroidal and cometary bombardments in its earliest history?

### 1.2.3 *Late accretion*

Late accretion refers to the sweep-up of leftover debris from the planet formation process after the Moon's formation and up to the present day. The most famous idea associated with the late accretion concept is the so-called Late Heavy Bombardment (LHB). This frequently invoked, but widely misunderstood concept, refers to a postulated hostile time in Solar System history from about 4.0 to 3.8 Ga in which the large lunar basins with known ages were produced (60). The LHB scenario is sometimes used to claim that life's start was delayed by a rain of large and devastating impactors to the surfaces of the terrestrial planets over a relatively short interval of time. Whether the LHB actually exists has been long debated: one view is that the basin-forming epoch marked the end of a steadily declining bombardment due to leftover remnants of planet accretion (61), while another proposes that the LHB was a short-lived "cataclysm" of dramatically increased impact rates onto the Moon (62,63). It is often stated, again without evidence, that the LHB had the capacity to repeatedly sterilize Earth up until around 3.8 billion years ago (64), and thus the origin of life had to have come well after the bombardment ceased, perhaps around 300 Myr later (65). However, re-analysis of radiogenic isotope ages (66) – variably sensitive to different temperature regimes – for asteroidal meteorites, coupled with models of the thermal fields of impact bombardments to the terrestrial planets (67), show that the abating impact flux from late accretion is best explained by a smooth monotonic decline in impactors rather than being punctuated by intense spikes. The impact frustration argument (68) in general (and the LHB mechanism for early life's repeated demise in particular) has fallen out of favour owing to thermal-physical models that describe a relatively benign thermal environment of the Earth at any time in late accretion bombardment since the last colossal impacts that affected Earth which gave rise to the Moon and the late-veneer pollution of Earth's mantle by highly siderophile elements like Pt, Pd, and Ir (29,45,69,70).

The output of these various models show that the cooling time to nominal conditions (290K) between the most destructive large impacts (asteroids >100 km diameter) from late accretion was hundreds to thousands of times shorter than the recurrence interval of individual large impacts (71). Hence, the impact flux was inadequate to sterilize the surface zone. Perhaps it comes as no surprise that different lines of evidence for a biosphere at ca. 3.8 Ga from characteristic stable isotope fractionation patterns of Carbon (72-77), Iron (78), Nitrogen (79) and Sulfur (80,81) seem to show that life has existed on Earth since the late Hadean, perhaps emerging within the first ~200 Myr of the solar system (82). It is interesting to speculate about what this first life could have been like.

### **1.2.4** *Early Earth vs. the RNA World*

An early waypoint in life's emergence from the non-living to the living may have been to use ribonucleic acid (RNA) in both genetic and catalytic roles in what has been dubbed the "RNA World" (83) (**Figure 6**). This early transitional history of cellular life – not including viruses – probably occurred before the contemporary role of de-oxyribonucleic acid (DNA) was established. Although many alternative hypotheses have been presented that begin with life as a suite of increasingly complex metabolic cycles driven on mineral substrates (84), or as a structural entity that had encapsulated chemistry different from its surroundings (85), that RNA may have come first is supported by a number of interesting lines of evidence. Among these are that the catalysis displayed by ribosomes in all three kingdoms of Earth life (86). Furthermore, self-splicing introns (87), many primal non-informational RNA molecules (88), along with the intermediate nature of messenger RNA in the biosynthesis of proteins in biology, as well as RNA cofactors found in metabolism throughout the biosphere (89), attest to the antiquity of RNA. The "hard" RNA World hypothesis is also used as a model for the origin of life on Earth and terrestrial planets in general, such as Mars in our Solar System (16,90; cf. 91). This RNA First idea purports that natural geologic conditions gave rise to oligomeric RNA with information storage (>100 bits, <10k bits; 92) adequate to support reproduction and thus Darwinian evolution. Yet, the molecular structure of RNA, it has been argued, is far too complex for it to have emerged by itself under natural conditions (93,94). Nevertheless, laboratory experiments are ongoing that seek to generate oligomeric RNA from monomeric ribonucleic acid structures in environments that were present on the Hadean Earth (95,96).

<INSERT FIGURE 6 HERE>

The ribonucleic acid structures of the RNA World, however, were probably not very stable in most planetary surface environments, which restricts the possible space for them to have emerged in the first place. They break down with heat (>50˚C) and in the presence of common cations in seawater: $Mg^{2+}$, $Ca^{2+}$, $Mn^{2+}$, $Fe^{2+}$. Maybe the first steps of the RNA World were confined to cool, lacustrine systems such as endorheic lakes (97). Furthermore, the reactions

RNA can catalyze are limited and it has to have some external source of nucleotides. Everyone, including the most ardent “RNA World First” enthusiast, agrees that these organisms eventually given rise to the DNA/Protein world to overcome these structural disadvantages. If the *Lavosier-Lomonosov Bridge* included an RNA biome as a transitional form inhabiting relatively mild aqueous near- or at-surface regions (98-101 see ref. 102), they were likely flimsy proto-organisms (103) eminently susceptible to extinction from successive regional thermal stresses to the hydrosphere from the largest impacts from late accretion. In contrast, DNA is much less sensitive to heat than RNA, and is stable in the presence of common marine cations. Furthermore, proteins with 20 amino acids can catalyze an enormous variety of chemical reactions that would otherwise be unavailable to an RNA organism, and peptides support and repair thermal damage to DNA. It was previously speculated (67,69) that the successors to the RNA World were thermally more robust RNA-peptide and then RNA-DNA-peptide life forms that adapted to colonize deep and hot crustal settings near hydrothermal vents (81). This argument merely makes the assertion that wherever life originated, the early history of biology on our planet includes exclusive residence by deep hydrothermal vents, it does not imply that there is where life necessarily began.

Why are hydrothermal environments so important in any discussion of early life? The crustal regimes of hydrothermal vents exist wherever hot rock and liquid water meet, whether in or on planets, moons or asteroids. These milieux exist at the interfaces of the reduced (electron-rich) interiors/mantles, and the oxidized (electron-poor) exteriors/surface+atmosphere. On a planet like Earth, such environments are volumetrically enormous in the vast plumbing network of the oceanic crust and on continents (and within them, in deep sedimentary basins), and where they are shielded from impacts and the intense ultraviolet radiation output of young Sun-like stars (104,105). Perhaps it is a general case that such hydrothermal refugia for early life in planetary crusts allows RNA-DNA-peptide biospheres to ride out the waning stages of bombardment in young planetary systems. If so, these conditions lead to the DNA-RNA-protein World going on to repopulate terrestrial-type planets from the inside→out.

Now that we have set the guideposts for life in planetary-scale environments, we can begin to consider planets around other stars and whether or if they are able to satisfy the criteria fulfilled by the Earth system.

## 2. *The Lavosier-Lomonosov Bridge* on small rocky planets around other stars

To understand the prospects for life as we know it on Earth-like exoplanets, it makes sense to take stock of what these worlds are and what they could plausibly be made of (106). Since the first verifiable discoveries of exoplanets around main-sequence stars in the 1990s: 51 Pegasi by Mayor and Queloz in 1995 (107) and HD 114762 by Latham and co-workers in 1989 (108) and, more than 4500 planets have been confirmed as of December 2021 (https://exoplanetarchive.ipac.caltech.edu/) with nearly 1000 in the radius range ≤1.25 $R_\oplus$ that could qualify them as in the “terrestrial-type planet” category (109). Altogether, the suite of exoplanet finds runs the gamut of those smaller than Mercury to many times larger than Jupiter (**Figure** 7).

<INSERT FIGURE 7 HERE>

Five lines of evidence are commonly present in our data set of exoplanet observations that could bear on the potential for life on worlds beyond our Solar System:

- (i) location around the star;
- (ii) planet mass;
- (iii) planet radius;
- (iv) atmosphere and clouds (very small data set); and,
- (v) age of the host star and thus a value for the minimum age of planets that orbit it.

Exceptions to this list would be a tidally-heated exomoon of terrestrial planet mass around a Jovian-type planet with the potential to host conditions for life well outside the nominal habitable zone that does not only rely on insolation from the host star (e.g. 110). Another is that in some extremely rare cases a rogue planet can be captured by a star, and thus may be of any age (older) and different composition (111). Planet capture is, however, dynamically very difficult, and in any case such captured irregular planets are going to be far away from their primary in wide orbits.

### 2.1 Modelling terrestrial-type exoplanet compositions

At the present state of technology, almost all inferences about an exoplanet's possible surface (atmospheric) and geophysical (crust/mantle/core) regimes require modelling, which itself necessitates extrapolations and presumptions based on knowledge of our Solar System plus whatever other lines of evidence may be available. Numerous attempts over the last decade have been made to model the interiors of the burgeoning terrestrial exoplanet inventory. A suite of papers have explored thermal histories, and the plausibility of present geological activity (e.g. 112-129; see ref. 15 and references therein). Many of these studies determined the theoretical tectonic regimes of a special class of exoplanets, termed "super-Earths" (>1-<10 $M_\oplus$; ref. 119). Such works more often than not mention "life" and/or "habitability" as a primary or secondary motivations, but the first consideration needs to be on the nature of the planet itself before we can begin to consider whether it could have the capacity to host life.

#### 2.1.1 *Planetary diversity*

Several works (109,130,131) proposed a radius cut-off at ~1.5 $r_\oplus$ between what are known as "super-Earths", and a class of planets dubbed "mini- (or sub-) Neptunes" at ~2 $r_\oplus$. These are plausible upper limits for planets that are broadly Earth-like and thus capable of hosting a biosphere of the kind we are familiar with, and thus are worth adopting as mass limits.

Like the elements of life, planets are formed from common nuclides in the galaxy. Comparisons of elemental abundances in the Solar System, and spectroscopic of average Planetary Nebulae[8] and the Orion Nebula – assumed to represent the present composition of the ISM – show that the top 9 elements (normalized to H at 12.00; not including noble gases) are: C (8.6), N (8.0), O (8.8), S (7.2), Ca (6.4), Na (6.3), Cl (5.5), K (5.5) and F (4.6) (ref. 132). Other major nuclides present in the ISM are Si, P, Fe, Co and Ni, but these are difficult to measure because of scavenging by grain formation in molecular clouds (133) The compositional differences between Solar System, Planetary Nebulae and ISM are probably small, but as we shall see (**sec. 3**) are significant in that small differences in elemental ratios can result in different planetary states. With this in mind, it is probably also the case that "exotic" planets with hundreds to thousands of times enrichment in, for example, heat-producing actinides like U and Th, relative to Solar System values are difficult to reconcile with actual abundance observations (cf. 134).

#### 2.1.2 *Planetary internal heat production*

Heat production within a planet's mantle, which on Earth manifests itself via plume activity, plate tectonics and crustal structure and composition, is determined by the long-lived, heat-producing radionuclides created during earlier stellar nucleosyntheses that get injected into the interstellar medium (ISM) from which all stars and their planetary systems form. Long-lived radioisotopes dictate Earth's modern heat budget, complemented by lingering heat from accretion and differentiation. Heat early on was also provided by short-lived nuclides, principally $^{26}$Al with a very small (1% relative) contribution from $^{60}$Fe (135). Earth's own heat production has changed markedly over geologic time (136) (**Figure 8**). Every planet in our Solar System, including Earth has its complement of $^{235}$U that is 88× less than at time of Solar System formation ca. 4.57 Gyr (137,138). Significantly, the ability of a planet to sustain a mantle convective regime and plate tectonics changes with its evolving thermal profile due to radioactive decay over geologic timescales (123,139). Nevertheless, up until recently exoplanet geodynamic modellers were compelled due to lack of better constraints to assume modern Earth, primordial Earth, or chondritic values for the concentrations of radionuclides. Indeed, many exoplanet models are forced include heat production rates at steady-state with Earth values instead of those that decline with time. It is worth emphasizing at this point that nothing makes sense in geology, exoplanetary or otherwise, except in the context of geologic time.

[8] Planetary nebulae are expanding shells of luminous gas ejected by dying stars; about 20,000 planetary nebulae are thought to exist in the Milky Way Galaxy.

<INSERT FIGURE 8 HERE>

Owing to its relatively rapid decay rate, the shortest-lived of the principal heat-producing nuclides operative over geologic time, $^{235}U$, became a subordinate heat source in Earth's mantle after less than ~2 Gyr of our planet's history. This has left the other three heat-producing isotopes to sustain the heat, along with leftover heat of accretion, that helps drive Earth's present geological activity. Over the last ca. 4.6 billion years and with a half-life comparable to the age of the universe, $^{232}Th$ has lost a mere 20% of its original abundance, whereas $^{40}K$ has lost ~90%. When Earth is 10 Gyr old, its total radiogenic heat production will be ~15% of that at time of formation 4.5 billion years ago. By then, neither $^{40}K$ nor $^{238}U$ will heat contributors, and $^{232}Th$ will go on to weakly power the interior. Approximately 900 Myr from now, there may no longer be enough heat in the mantle to sustain Earth's mobile-lid convection, in which case it is predicted that plate tectonics will cease (140). Radiogenic heating is therefore, as a sub-category of bulk composition, a pivotal parameter to the geology of any rocky planet. Data from GCE-coupled thermal models (141) suggest that the thermal window for a geologically active cosmochemically Earth-like planet is about 6 Gyr. After that, small rocky planets become thermally moribund and the planetary redox couple at the interface of the crust and atmosphere tends towards equilibrium. Equilibrium is synonymous with dilapidation and from the point of view of a biosphere, death. The smaller the planet, the shorter the interval of time between its formation this equilibrium state unless an external forcing mechanism allows for sustained heat production to facilitate melting of rock (such as in a tidally-heated exomoon of Mars' mass or greater orbiting a SuperJupiter; ref. 142).

### **2.2** *Composition of the star*

Our closest star, the Sun, is 99.8% of the mass of the Solar System and by definition represents the average composition of the whole system. The proportions of, specifically, the rock-forming elements (e.g. O, Si, Al, Fe, Ca, Na, K, Mg, etc.) available in the interstellar medium and the molecular could from which stars and planets form, ultimately determine the available components for building the planets (143). The geophysical foundations of terrestrial planets rests on the nucleosynthesis of rock-forming elements in the Galaxy over billions of years as recorded in the composition of the Sun. Thus, by knowing the Sun's composition we have clues on the compositions of the planets. Nucleosynthesis studies that seek to reproduce the composition of the Sun and of other stars and the planets around them, however, must have an accurate, physically motivated explanation for the pattern of abundances that are observed. The sources of this information come from the Solar System itself (i.e., the Sun) and in other locations in the cosmos (other stars, the ISM, cosmic rays, inter-galactic medium and other galaxies; Magellanic Clouds). The key to that understanding is accurate information on the pattern of these abundances.

For solar abundances there are three main sources: bulk compositions of Earth, Mars, the Moon, asteroid Vesta and other asteroidal meteorites; bulk compositions of carbonaceous meteorites; and, of course, the Sun itself, mostly in the solar spectrum (144). This subject has a long history. The first person to (perhaps unwittingly) explore the Sun's composition was W.H. Wollaston, who was inspired by Newton's experiments with light and prisms from 1666 (145). He was the first to publish a description of the distinctive patterns of bright and dark lines in the prismatic spectra of the Sun and various other sources including electricity and candlelight (146). Soon after, J. von Fraunhofer read of Wollaston's discoveries, and improved upon the purity and clarity of large prisms to invent the precursor to the modern spectroscope. In the course of his experiments with firelight from a furnace, in 1814-1815, Frauenhofer discovered a bright fixed orange line which enabled him afterward to determine the absolute power of refraction in different substances. Further experiments led to his discovery of more than 500 dark fixed lines in the solar spectrum. Millions of such fixed absorption lines have since been dubbed Frauenhofer Lines. In a true stroke of genius, Fraunhofer directed his attention to the sky and also detected dark lines in the spectra of several bright stars, including the A0/A1 star Sirius, but could show that they were in slightly different arrangements from the solar pattern (147). He concluded that different stars have different properties. The "Frauenhofer lines" were later explained by Kirchhoff and Bunsen (1859) as atomic absorption lines from diverse compositions of substances that were made incandescent (usually by flame). A major step forward in the planetary-solar chemistry connection came later with the work of F.W. Clarke (1889) who presented the relative abundances of chemical elements for different Earth reservoirs (crust, ocean), and as an early proponent of Mendeleev's classification of elements in the Periodic Table, used this to organize his data. This work naturally led to the conclusion that the cosmic abundances of the elements in the Sun are reflected in the composition of the most primitive rocks that fall from space (meteorites), and which follow the Sun's values as determined by Frauenhofer absorption spectra (148). As such, reveal something about Earth's bulk chemical makeup of the rock-forming elements (149-151). In discussing the origin of life on Earth, J.D. Bernal in 1947 (152) mentioned the use of stellar abundance measurements of stars to compute the composition of primitive Earth's crust, and thence of the minerals that provided the *Lavosier-Lomonosov Bridge* to the animate world of biochemistry. Once compilations were made of stellar photosphere abundances against the CI (Ivuna) group of relatively primitive carbonaceous chondrite meteorites, it became clear that a near 1:1 correspondence between the composition of the Sun and of various meteorites exists for many elements (153,154). It has since been found that over forty elements have a Si-normalized photospheric/CI chondrite abundance ratio between 0.9 and 1.1 (155,156). To first order, the bulk composition of Earth and of Earth's mantle is the average Solar System composition (157) for the rock-forming elements (excluding volatiles such as H, N, C, O, and Li, as well as the noble gases) in the Sun and carbonaceous meteorites like CI (**Figure 9**). The deviation of solar values in volatile elements with Earth values, however, is large enough that it causes problems for quantitative derivation of Earth's composition from direct stellar spectroscopy values. We return to this problem, later. Meanwhile, it was actually by linking cosmic abundances of the elements and isotopes to primitive meteorites that Suess and Urey in 1956 (158) and Cameron in 1957 (159) presented complementary cosmo-geochemical arguments which firmly placed testable foundations to the physics behind nucleosynthesis theory published a year later in $B^2FH$ in 1957 (160). If the Lomonosov-Lavosier Bridge is universal, then we need to understand how the elements are distributed in planets in the Galaxy.

<INSERT FIGURE 9 HERE>

### **2.2.1** *Cosmochemical inheritance*

Of the many profound conclusions in $B^2FH$, one of the most important is that all but the five lightest elements owe their origins to synthesis inside of stars (**Figure 10**); the first three ($_1$H, $_2$He, $_2$Li) formed in the Big Bang, and the next two ($_4$Be and $_5$B) by cosmic ray interactions with interstellar atoms (14). The rest are created in various stellar nucleosynthetical processes and mixed into the ISM at the end of a star's lifetime. The mixing occurs on timescales shorter than the half-lives of the radioactive nuclides that power terrestrial-type planet interiors. Relevant radiogenic isotopes driving heat production, with half-lives ($\tau$1/2), are: ($^{40}$K, $\tau$1/2=1.25 Gyr; $^{232}$Th, $\tau$1/2=14.05 Gyr; $^{235}$U, $\tau$1/2=0.704 Gyr; and $^{238}$U, $\tau$1/2=4.468 Gyr). The galactic spiral period is ~220-360 Myr (161,162) whereas the galactic period in the solar annulus is 250 Myr, which is a factor of ~2 shorter than the half-life of $^{235}$U. This observation is important because the products of GCE mix in rapidly and we can say that by knowing the evolution of stellar nucleosynthesis of the planet-forming elements and the ages of different stars that host planets, we are able to broadly deduce the evolution of planetary compositions through galactic time.

<INSERT FIGURE 10 HERE>

## 2.3 Chemical evolution of the galaxy and planets

Several implicit treatments, have, over the years, explored the time-dependent linkage between GCE models and planetary composition (163-167). Despite the fact that more GCE models are becoming available for this kind of work (168,169), a common shortcoming in such models is the adoption of the composition of the present Earth as a standard in composition that is simply scaled up or down in size. Based on variations in composition of their host stars, however, it is much more likely that some or even most exoplanets could be substantially different in their bulk (major-element) and trace-element composition from any planet in our solar system. Owing to this, the effects of parameters such as mass, differentiated nature, water content, and even orbital dynamics (170) often yield contradictory results because determining factors are extrapolated or simply assumed based on "chondritic values" or "Bulk Earth". Although the first application of radioactivity to estimate the age of the cosmos was proposed by Rutherford in 1929 (see ref. 171), it has only been very recently that GCE-based modeling of radioactive nuclide production has been specifically turned to the study of exoplanets and long-lived heat production questions (141,172). A summary of the key aspects of the approach by Frank and co-workers (141) which combines GCE with cosmochemistry to understand the elemental feedstocks to stars and planets in the galaxy over time, follows.

### **2.3.1** *Elemental feedstocks to molecular clouds that form star-planet systems*

Galactic chemical evolution models are formulated to address how the bulk chemistry of the galaxy changes in both time and space as old stars perish and new generations arise (e.g. 173). All GCE models share four common components: (i) boundary conditions, such as the galaxy's initial composition; (ii) stellar yields of heavy nuclides; (iii) star formation rate (SFR) and initial mass function (IMF); and (iv) gas inflows/outflows to the galactic system (174). Quantitative constraints to GCE models are provided by solar and meteorite compositions, abundance ratios of the isotopes in primitive meteorites, metallicity of G-dwarf stars measured from spectra, and galactic abundance gradients (175). Interestingly, the effectively instantaneous appearance of heavy elements (including r-process) in the earliest galaxy as seen in the oldest halo stars was due to production in the first generation of massive stars that lived ~$10^6$ years after galactic formation (176,177). Further, the discovery of extreme star formation rates, or EFR (>100 that of our Milky Way) in a galaxy (z8_GND-5296) which formed ca. 700 Myr after the Big Bang (178) is evidence for the kind of nucleosynthetic environment that led to rapid early enrichment of r-process nuclides even in the early universe. The long-lived radioactivities which are synthesized from O-burning, s-, and r-process start to decay as soon as they are generated. Although it is unclear how the supernova rate or neutron star merger rate has changed over time, actinide abundances for the early Solar System are roughly that expected for near-uniform production (179,180).

Overall, computational GCE models, foremost being the Clayton Model (181-183), parameterize galactic infall in such a way that they provide analytical solutions for disk gas mass, total solar mass, and metallicity. This means that such models include many assumptions. One of these is the instantaneous recycling approximation (IRA), which states that stellar lifetimes are negligible relative to the timescale of gas consumption, and another the instantaneous mixing approximation (IMA), which assumes that stellar ejecta are mixed instantaneously into the ISM. The initial mass function (IMF), which sets the population distribution of stars, and the SFR is set to depend linearly on the gas mass. However, it is recognized that SFR is discontinuous, the production of short-lived nuclides is patchy, and delays exist due to stellar lifetimes and inhomogenous mixing in star-forming regions (184). Thus, GCE modelling of the evolution of the rock-forming elements requires further development to account for these parameters and thus to make better sense of astrophysical observations (185-188). An important aspect to any GCE model is that it tracks the evolution of the total mass (relative to the initial mass) of the solar annular region of the galaxy, the mass in gas, and the mass in stars (**Figure 11**). This "mass" is the elemental feedstock of the molecular clouds from which stars and planets form.

<INSERT FIGURE 11 HERE>

There are other assumptions with analytical GCE models which arise due to their engineered simplicity. After all, they attempt to tackle a system (the Milky Way galaxy) that is ~$10^5$ ly across and 10+ billion years old. One of these is that mass of the solar annulus builds up due to infall, but this rate declines with time so the rate of growth of the total gas mass not tied up in stars also declines; the Frank model is tuned so that over time it reaches the present ~20% value gas:star value (189). The choice of model parameters used by Frank et al. (141) was further calibrated using two key observations. First, it was designed to be broadly consistent with the G dwarf star age-metallicity relationship (182). Second, Madau and co-workers. (190) showed from observations of distant galaxies that the SFR peaked at redshift of ~1.5, corresponding to a time between 2-3 Gyr after the Big Bang. The peak in gas mass, and by proportionality of this quantity to the SFR, is at about 4 Gyr. The gas mass for the galaxy as a whole is ~0.1 to 0.15, which would lower the gas fraction and require a larger rate of gas consumption from star formation (term ω in the model of ref. 141).

### 2.3.2 Synthesis and concentration evolution of the rock-forming elements

Mirroring to no small degree, the most common rock-forming elements of the bulk Earth are (ref. 191): Fe (31.9%), O (29.7%), Si (16.1%), Mg (15.4%), Ca (1.71 %), Al (1.59%), Ni (1.82%), and Na (0.18%). Nucleosynthetic output from the Frank model for the major elements is shown in the constant enrichment in the ISM of the stable, primary-only nuclides Mg, Al, Si, and Ca in **Figure 12**.

<INSERT FIGURE 12 HERE>

The second tier of elements that comprise the bulk Earth (H, C, K, S, P, Cr, Mn, Co) are all individually less than (0.6%), and the remainder are at low concentrations. To reproduce these concentrations, the corresponding yields in terms of mass fractions in the Frank model are fit to Solar System values at t = 8 Gyr (if the Solar System formed 8 Gyr into galactic history; 193). For example, the model provides a reasonable match – considering the many variables that exist – to the overall declining trend of Mg/Si in stars across a range of metallicities [Fe/H] equivalent to, or slightly less, than the solar value (**Figure 13**). The Frank model treats the co-evolution of [Mg/Si]/[Fe/H][9] as a function of time and assumes Fe/H increases with decreasing stellar age; the comparative database of stars used in this particular analysis taken from Adibekyan and co-workers (194) is not similarly plotted with age and therefore discrepancies are anticipated between the two trends. We now analyse the case of iron.

<INSERT FIGURE 13 HERE>

### 2.3.3 *56-Iron evolution in the galaxy*

The nuclide with the third highest binding energy per nucleon is $^{56}$Fe (after $^{62}$Ni and $^{58}$Fe; ref. 195). It is, however the most abundant isotope of iron (91.75%), and furthermore based on planetary and meteoritic samples Fe cosmochemically exceeds Ni abundance by a factor of ~17 (ref. 196). The origin of $^{56}$Fe is a special case in nucleosynthetic production models due to an additional source found in Type 1a supernovae (197). Type 1a supernovae began to contribute to the ISM around 1 Gyr after galaxy formation (198). From then, $^{56}$Fe production should have gradually increased. This is important because not only are Fe and Ni the principle elements of metallic core, there is also an apparent relationship between metallicity – expressed as Fe abundance in [Fe/H] and the presence (or absence) of giant planets (199) (**Figure 14**).

<INSERT FIGURE 14 HERE>

This can be explored temporally by modelling the production of $^{56}$Fe. Model yields of $^{56}$Fe can be made by summing two primary terms: α1 due to synthesis in massive stars and α2 due to

[9] The units represent orders of magnitude values, or, as used by astronomers, 'dex'. The value converts the number before it into its base-10 antilogarithm so that 1 dex is $10^1$ = 10 (see ref. 192).

Type 1a supernovae. Early in galactic history, we can assume the yield to be pure α1, but as the galaxy ages, the yield should evolve to α1 + α2. Then, we can choose α1 and α2 and the timescale for the evolution of the total $^{56}$Fe from α1 to α1 + α2 to reproduce the observed galactic disk [O/Fe] vs. [Fe/H] trends (200). As a proof-of-concept, a preliminary model of $^{56}$Fe production and enrichment in the ISM by this formulation is shown in **Figure 15**. The next nuclide to consider is potassium.

<INSERT FIGURE 15 HERE>

### **2.3.4** *The special case of $^{40}$K*

Potassium-40 is the principle heat-producing nuclide of Earth's mantle in its first 2.5 Gyr. It is a difficult case in that it can be produced as both a primary and secondary nuclide species (14). While its abundance in the Solar System at time of formation is known from back-calculation from present values in primitive meteorites, the relative proportion of the two nucleosynthetic contributions is not and currently a topic of intense study (201). Detailed $^{40}$K yield data from modern stellar models (202) aim to constrain the primary vs. secondary aspect of this isotope. Owing to this uncertainty, models await formulation of the end-member scenarios in which $^{40}$K is only primary or only secondary, and investigations of proxy nuclides for $^{40}$K abundances in stellar spectra. Best estimates based on simple assumptions for $^{40}$K abundance in the ISM with time are provided in **Figure 16**.

<INSERT FIGURE 16 HERE>

## **2.4 Lingering problems**

The age of the galactic disk and the solar annulus within it, is not necessarily the same as the computed ca. 12.5 Gyr galactic "age" (cf. 193). This is essentially the age of the globular clusters, galactic halo and perhaps part of the bulge. Other age estimates for the galactic disk/solar annulus of ca. 9.5 Gyr have been offered (203,204). The Frank et al. code (141) that we discuss the most in the present work used the Dauphas (193) timescale of 12.5 Gyr. New models are warranted to explore) (more likely) shorter galactic timescales such as 9 or 10 Gyr.

### **2.4.1** *GCE helps us understand the compositions of planets, not define them*

Even though the analytic model described here provides the essential framework, much more detailed multi-dimensional GCE models are required to move the field forward. For example, general purpose multi-dimensional isotope mixing codes are currently being developed that follow the mixing and transmutations of species in an arbitrary number of (and arbitrarily linked) zones. Chemical evolution modules such as these follow the evolution of planet-building elements in a multi-dimensional ISM, that also include the effect of stellar lifetimes. While far more computationally intensive than analytic models, multi-dimensional, time-dependent calculations would allow us to relax the two key (and probably incorrect, see ref. 188 and references therein) assumptions of instantaneous recycling (IRA) and instantaneous mixing (IMA) that are otherwise widely assumed. For example, in the simplest analytic models, Mg and Si are both dominantly labelled as primary nuclides, and since the analytic models do not consider individual stars but rather generations of stars, the [Mg/Si] ratio changes little during galactic evolution compared to the real scatter present in stars. In reality different stars have different yields of Mg and Si. This means the local [Mg/Si] can vary in time and space (**Figure 17**). This point has important implications for geologies of exoplanets that we explore in **Sec. 3**.

<INSERT FIGURE 17 HERE>

The Frank et al. (141) model made simple assumptions not just about the production of elements through galactic time, but also with scaling the outputs of the GCE model to terrestrial exoplanets. These assumptions are justified for conceptual analyses, but are arguably too inexact, and must be combined with direct observations of planet-hosting stars from spectroscopic studies, guided by geochemistry (**Sec. 3.X**). This is because when planets form and differentiate, elements fractionate according to their geochemical affinities for core, mantle or crust (**Table 1**). Lithophile elements will end up primarily in the mantle and crust. Differentiation separates metal-loving (siderophile) elements from lithophile elements as metal descends into the core. From an early stage, the bulk silicate Earth (BSE) experienced this fractionation of incompatible elements into crustal components (reviewed in ref. 205).

**Table 1.** Geochemical behaviours and partitioning classifications* of the planet-forming nuclides (and water) and their 50% condensation temperatures from solar gas at partial pressure of 10.13 Pa ($10^{-4}$ atm). (154,206). Those elements discussed in the text of this paper one way or another, are shown in **bold**. Noble gases are not treated.

| **Category** | **Temperature range (K)** | **Lithophiles** | **Siderophiles** | **Chalcophiles** |
|---|---|---|---|---|
| Refractory | ≥1300 | Zr, Hf, Sc, Y, REEs, **Th**, **U**, Al, Ti, Ta, Nb, **Ca**, Sr, Ba, V, **Mg**, **Si**, Cr, (**Fe**), (**C**) | Re, Os, W, Ir, Mo, Ru, Pt, Rh, Ni, Co, (**Fe**), Pd | None |
| Moderately refractory | 1230-800 | **P**, Mn, **K**, Ga, **Na**, **Cl**,** Rb, Cs | Au | As, Cu, Ag, Sb, Ge |
| Highly volatile | 750-250 | **F**, **Cl**,** Br, I, (**C: $CH_4$, $CO_2$, HCN**) | None | Bi, Pb, Zn, Te, Sn, **Se**, **S**, Cd, In, Tl, Hg |
| "Ultra volatile" | <182 | **H/$H_2O$**, **N** (**$N_2$, $NH_3$**), (**C: CO**) | None | None |

*Geochemists classify the elements (and compounds, like $H_2O$, $N_2$, $CH_4$, $CO_2$, CO, etc.) into lithophile (silicate), siderophile (metal), and chalcophile (sulphide) based on partition-functions, chemical affinity, redox potential and valence state, and physical+mechanical properties. The classification is loose with some overlap (see Fe, C). Without $Fe^0$ or $Ni^0$ present, for example, many elements behave as lithophiles.

**Chlorine condensation temperature might be similar to that of Br and I (ref. 207).

### **2.4.2** *Making sense of likely vs. unlikely characteristics of rocky exoplanets*

The general lack of astronomically-constrained GCE models for exoplanetary systems severely hampers our understanding of what could be considered "likely" or "unlikely" in terms of plausible compositions. Several studies have noted a wide variation in [Mg/Si] and [C/O] for giant planet-hosting stars (109). Pioneering work trying to tie planetary compositions to stellar values in accretion simulations (e.g. 187) emphasized stellar Mg/Si and C/O as the strongest control over the bulk mineralogy of solid materials that participated in building exoplanets. This would be a good assumption if the data for C/O were more dependable (208). Therefore, owing to the fact that [C/O] data for stars are under debate, and show variation that can lead to models for "exotic/strange" planets such as pure diamond- or graphite-worlds (cf. 208-211, it makes sense to focus instead on nuclides that heat- and build silicate minerals in planetary mantles and crusts.

One of the most significant findings in exoplanet chemistry is the strong correlation between [Fe/H] and giant-planet frequency, which also follows all the rock-forming metals (212). The origin of this trend has been a matter of vigorous debate, with three main competing theories (213): (i) primordial metallicity of the stars with giant planets; (ii) pollution by post-formation accretion of gas-depleted material onto the host star; or (iii) selection effect from metal-rich stars with planets which are easier to detect by radial velocity. Current data support the primordial hypothesis (point i). It is worth mentioning in conclusion that those outlier systems believed to have undergone unusual source origins and formational histories should be considered in detail, individually. Considering the numerous assumptions required in GCE models and stochastic effects that determine final planetary composition, it is worth restating that GCE models are intended to guide analysis of plausible terrestrial exoplanet compositions, not define them. For that we must turn back to Frauenhofer and spectroscopic analysis of stars. With such compositional information, we can begin to build a geophysical picture of extrasolar rocky planets.

## 3. Geodynamical inferences

A planet's thermal evolutionary pathway is dictated in large part by its mass (214,215). Planets need to be large enough to retain internal heat over long times for continued geologic activity. Both the Earth's Moon (0.0123 $M_{\oplus}$) and Mars (0.107 $M_{\oplus}$) witnessed variably rapid early cooling as their accretionary heat dissipated and radiogenic heating failed to sustain vigorous internal convection. Internal heating also maintains a convective outer core that sustains a magnetic field, another potential consideration for "habitability" (216). Mars is observed to have crustal magnetic anomalies, suggestive of a past magnetic field (217). Knowing the relative proportion between radiogenic heating and secular cooling of a planet, even of our Earth, remains a challenge. This concept is quantified by what is termed the Urey ratio, which is the contribution of radiogenic heat output relative to the total heat budget of a planet (in the case of Earth, as measured through its surface). The Urey ratio is debated, with estimates ranging between 0.21 and 0.74 (ref. 218, and references therein). The prevailing view places the range of the contemporary Urey ratio from simulations (139) at 0.3-0.5, which is consistent with observational and geochemical constraints. Correcting for such factors as limited heating from the core and contribution of secular cooling, an estimate for Earth's present-day radiogenic heat production in the mantle is $7.38\times10^{-12}$ W $kg^{-1}$ (e.g. 219) (see **Figure 8**). The general rule is that the greater the mass of the planet, the more heat it will retain from accretion and core formation, and generate from intrinsic radioactivity. This rule can be violated in cases were planets have smaller cores and larger mantles (more radioactivity and therefore hotter interiors), or wildly different mantle compositions that lead to rheological conditions that mitigate heat loss owing to suppressed convection (again, yielding hotter interiors). Planets with large core:mantle ratios that happen to have vigorous convection will maintain relatively cooler interiors.

### 3.1 Iron availability

Iron content is also tied to the thermal evolution of silicate planets. Observations show that [Fe/H] has a weakly decreasing trend as a function of increasing stellar age (220). As the galaxy matures, stars and their planets forming now might be expected to be more enriched in Fe as they are in U, Th, and K, than when our Solar System formed at 4.57 Gyr (**Figure 18**). Owing to the delay in Fe nucleosynthesis from Type 1a SN, it may be the case that Earth-like rocky planets formed early in galactic history accreted with small Fe-Ni cores, leading to what has been termed a "super-Luna" planet (141). This viewpoint inverts that of Mordasini and co-workers (221) by applying the stellar [Fe/H] trend to plausible (and testable) interiors of mini-, eta- and super-Earths. An Earth-mass planet with a relatively small core like the Moon (R=1791 km, $R_{core}$= ~330 km; $R_c$/R=0.18) will have a correspondingly larger silicate mantle. The radionuclides Th, U and K are lithophile elements that reside in the mantle; they do not partition into the (metallic core). This means that planets with large mantles have the capacity have more internal heat than, for example, an Earth-mass planet with a core radius proportional to that of Mercury (R=2440 km, $R_{core}$= 2030 km; $R_c$/R=0.83) (see, for example, ref. 222).

<INSERT FIGURE 18 HERE>

### 3.2 Different (exo)planet compositions lead to different geophysical states

The mantle is 67.3% of Earth's mass, and 84% of its volume. Earth's core accounts for 32.5% of the planet's mass (15% by volume), and assuming that the core contains only Fe, Ni and S means that mass balance can be used to calculate the composition of the silicate mantle. This would be the composition of bulk silicate Earth (BSE), or "primitive mantle" (PM; basically, the composition before crust formation). Following Palme and O'Neill (157), we can construct a simple model of the mantle's composition if we assume only the six rock-forming elements listed above comprised the BSE. Along with the oxygen associated with them by stoichiometry – and with negligible amounts of $Fe^0$ and $Fe^{3+}$ in the BSE – the sum of their oxides must add up to 100 weight %, so that

$$MgO + SiO_2 + Al_2O_3 + CaO + FeO = 100.$$

If we insert the solar abundances translated into the oxides for these elements, we get:

$$MgO = 38.12 - FeO/2.623.$$

This equation is valid for the mantle of a planet from our Solar System with carbonaceous chondrite bulk composition. It also means that the maximum MgO content of a completely reduced, FeO-free mantle of an Earth-mass terrestrial-type planet with otherwise Earth-like (MgO) composition, is 38.12%. This number goes down slightly if minor elements are added (e.g. Na, K). Variable enrichment in refractory lithophile elements (**Table 1**) should also produce lower MgO contents, whereas with increasing FeO, the corresponding MgO would decrease. Taken together, the lowest achievable MgO content for a (CI composition planet) without a core is 24.3% MgO. This is interesting, because re-writing the equation in terms of element minus oxides, and assuming CI bulk Earth Fe/Mg, yields:

$$Fe_{mantle} = (44.37 \times (1- X_{core}) - Fe_{core} \times X_{core}) / (1.571 \times (1 - X_{core})) \text{ [}Fe_{mantle} \text{ and } Fe_{core} \text{ in wt\%].}$$

This equation is valid for any planet with a volatile-free CI bulk composition and a metal core. It predicts a composition for Earth's mantle that is in surprisingly close (but not exact) agreement with the composition of the Sun, with some finer scale differences in the more complex real-life that are addressed below.

The interesting outcome of this analysis is that, like envisioned by Goldschmidt it may be possible to calculate Earth's composition assuming direct solar element ratios among rock-forming elements in the bulk Earth. The calculated mantle composition in **Figure 19** is generally close to the mantle composition derived from samples of upper mantle rocks. This argument recapitulates the old premise that the bulk Earth has basically a chondritic bulk composition.

<INSERT FIGURE 19 HERE>

### **3.2.1** *Problems with direct scaling of solar to planetary abundances*

There are, however, some important differences between calculated mantle compositions by directly using the solar model and the actual mantle composition measured in mantle rocks. As pointed out by Wang and co-workers (223,224): (i) Some major element ratios such as Al/Mg and Ca/ are much higher than solar, reflecting a general enhancement of refractory elements in Earth during terrestrial planet formation; (ii) It is likely that Earth's core contains a few percent Si (225), which causes the Si/Mg ratio of Earth's mantle to deviate from that predicted from the Sun; (iii) Earth is strongly depleted in volatile elements compared to the Sun, among these is S. Hence, computed S contents for the bulk Earth are unrealistically high; and, (iv) the bulk Earth Fe/Mg ratio is about 0.10 above the CI chondritic ratio (O'Neill and Palme,2008). In order to mitigate these differences, another technique is required that acknowledges the thermal, physical, chemical and mechanical processing of elements and compounds in the planet-formation process. This technique, termed "devolatilization" more accurately estimates the composition of a terrestrial planet from chemical analysis of stellar spectra and tested on the Sun-Earth pair (226). With this new tool, we can now begin to quantitatively assess terrestrial-type exoplanets by remote spectra analyses of planet-hosting stars.

### **3.3 Bulk exoplanet compositions**

In order to calculate bulk exoplanetary mantle compositions and the initial chemical profiles of the exoplanetary mantle from the composition of the host star, several processes have to be accounted for. First, the "exo-chondritic" solids, from which planets accrete, condense from the protoplanetary disk and can chemically react with one another. The composition of the protoplanetary disk is assumed to be that of host star, but elements are systematically fractionated during condensation depending on their refractory vs. volatile behaviour (154,155). Second, the condensed "exo-chondritic" solids aggregate to form planetesimals, which further collide with each other to form proto-planets, and ultimately, planets. However, compositional fractionation during the stochastic process of N-body collisions and planet formation is subdued assuming a relatively narrow feeding zone for the exoplanet (227). N-body simulations of planetary accretion tailored to the solar system support narrow feeding zones of the terrestrials (228-230). Third, the energy release during accretion leads to planetary melting and segregation of a metallic Fe-Ni core from the rocky mantle. The partitioning of iron between the core and the mantle primarily depends on the availability of oxygen, and thus on the oxidation state of the mantle. In any case Fe/Si of the planetary mantle is usually significantly lower than that of the host star. Finally, widespread melting in the mantle and recrystallization of a magma ocean formed by wholesale melting of a rocky planet during primary accretion will lead to further compositional fractionation. The first crystals that are formed in the deep magma ocean tend to incorporate Si and Mg over Fe (and Na). Likewise, heat-producing elements will be strongly partitioned into the mantle. It is indeed critical to account for fractionation during all the above processes in future exoplanet work to link the observed stellar composition to a realistic initial condition for models of exoplanetary dynamics.

### 3.3.1 *Quantitative abundances from devolatilization*

The formation of rocky planets like the Earth can be explained by the devolatilization i.e. depletion of volatiles – of the solar nebula, through processes such as evaporation, collisions and other thermal events (231,232). This feature of volatile depletion has been previously investigated and is called the "volatility trend" of the Earth in the literature (233). An example is shown in **Figure 20**, which compares the elemental compositions between the bulk silicate Earth (i.e. the bulk Earth excluding the core) and CI chondrites (the most "primitive" meteorites, often used as a proxy for solar abundances), as a function of the 50% condensation temperature, $T_C$. The chemical compositions of Earth, Moon, Mars, asteroidal meteorites and other rocky objects of the Solar System are broadly similar, which bolsters the old idea that the chemical composition of Solar System as a whole, and that of the silicate+metal planets in particular, generally reflects that of their host stars for refractory elements. This relationship, however, does not hold for volatile elements.

<INSERT FIGURE 20 HERE>

Consider the case of Na and K, which are more volatile than Fe/Si/Mg. In **Figure 21** it is obvious that the abundances of Na and K are well below that of solar and CI chondrite values. This is due to their volatility which is predictable by their chemistry. A characteristic feature of the comparison of protosolar abundances (i.e. based on photospheric and CI chondrite abundances) and terrestrial abundances, is the depletion of terrestrial abundances for elements with moderate condensation temperatures between 500 K and 1400 K. The depletion is systematic: the lower the condensation temperature, the greater the depletion. This depletion provides quantitative insights into the processes active in the early solar system and the fractionation of elements between gas and solid phases. Melting and vaporization experiments (232) and isotopic analyses (234,235) also yield complementary insights into devolatilization processes (**Figure 22**). With bulk compositional estimates for rocky exoplanets within reach, we can begin to consider what this means for their overal dynamical natures.

<INSERT FIGURE 21 HERE>

<INSERT FIGURE 22 HERE>

## 3.4. Bulk chemistry and dynamics

Bulk (major-element) composition determines which minerals are stable in a rocky planet's mantle. For low Mg/Si, substantial amounts of free $SiO_2$ are stable, along with $(Mg,Fe)SiO_3$ pyroxene (and high-pressure polymorphs). For Mg/Si≈1.0, the mantle mineralogy is strongly dominated by pyroxene-type minerals. For high Mg/Si, instead, mantle rocks are mostly (or even near-completely) made up of $(Mg,Fe)SiO_4$ olivine-type minerals (see **Figure 17**). The Mg#=Mg/(Mg+Fe) of olivine- and pyroxene-type minerals is strongly dependent on the availability of Fe in the planetary mantle. This availability does not only depend on bulk planet composition, e.g. in terms of Fe/Si, but is moreover controlled by the process of core formation, during which Fe is partitioned between the mantle and core. The bulk mineralogy and Mg# of the planetary mantle determine planetary dynamics and evolution through their large effects on physical properties and melting behaviour of rocks.

### **3.4.1** *(Exo)mantle viscosity*

The most important physical property for planetary interior dynamics is viscosity, which can differ by orders of magnitude between different minerals. Earth's upper mantle (not Primitive Mantle) with a Mg/Si≈1.3 is dominated (~60%) by olivine, and also contains ~40% pyroxene. As the viscosity of olivine is much lower than that of pyroxene (236), deformation in the Earth's upper mantle is governed by olivine rheology, and pyroxene remains mostly undeformed. In turn, the upper-mantle rheology of an Earth-like planet with Mg/Si≤1.0 would be dominated by pyroxene, and the relevant viscosity would be much higher, with important dynamical implications. Compared to the upper mantle, the Earth's lower mantle contains different minerals and is dominated by ~80% bridgmanite perovskite (i.e., the high-pressure polymorph of pyroxene) and ~20% (Mg,Fe)O magnesiowüstite (i.e., olivine-type minerals decompose into bridgmanite and magnesiowüstite under lower-mantle conditions). In this case, magnesiowüstite is several orders of magnitude structurally weaker than bridgmanite (237), thus dominating the rheology (and thereby viscosity) even at low volume fractions. Similar to the case above, the lower-mantle rheology of an Earth-like planet with Mg/Si≤1.0 would be dominated by bridgmanite, and hence the relevant viscosity would be substantially higher. The transition between the weak-mantle regime at Mg/Si>1.0 and the strong-mantle regime at Mg/Si≤1.0 is expected to occur across a narrow range of Mg/Si, since small volume fractions of a weak phase are sufficient to form an interconnected network, and govern the rheology (238).

An increase in mantle viscosity, such as due to planet compositions with Mg/Si≤1.0, is expected to result in more sluggish convection and delayed planetary cooling, compared to planets with an Earth-like Mg/Si. Thereby, planets with lower Mg/Si are expected to be hotter in their interiors, and undergo more partial melting near the surface. A decrease in viscosity results in a cooler planet because these are able to effectively conduct heat away. In turn, this affects the likelihood of the planet developing plate tectonics. Partial melting of rocks in the mantle results in magma that is typically less dense and thus rises and solidifies, forming crust that is also less dense than the parent rock. However, differences in bulk composition would change the composition of the magma and the resulting density of the crust. This is an area of research in exoplanet study that is ripe for study.

### **3.4.2** *(Exo)mantle oxygen fugacity*

Linked with the sensitivity of viscosity to composition, is the oxygen fugacity of the mantle system. With data available from devolatilization spectral analysis of stars, we can test the hypothesis earlier proposed by Ringwood (227) and described in Harrison (51) that posits:

(i) Earth's Primitive Mantle (not "upper mantle") value of [Mg/Si] is ~1.03 (CI=0.93);

(ii) As mentioned (**Sec. 3.4.1**) the dominant upper mantle (UM) phase of that composition is olivine $(Mg,Fe)_2(SiO_4)$, but this mineral has no lattice site that can accommodate ferric iron ($Fe^{3+}$);

(iii) If Earth had inherited a slightly lower [Mg/Si] (e.g. <1.0), pyroxene ($(XY(Si,Al)_2O_6$, where X represents divalent Ca, Mg and Fe, and Y represents trivalent Cr, Al, Fe) would dominate. Pyroxene takes up $Fe^{3+}$ into its structure and with substitutions maintains low activity of $Fe^{3+}$ and a very low oxygen fugacity;

(iv) Owing to (ii), the $Fe^{3+}$ present in the Earth's UM goes into spinel ($(Mg,Fe)Al_2O_4$) such that there is a modal phase which imposes a high oxygen fugacity (~FMQ) on gases in equilibrium with rock.

It seems that our planet's UM always degassed a relatively oxidized form of carbon ($CO$-$CO_2$) rather than an alternative mantle which would degas $CH_4$-$CO$ (see **Figure 5**).

Thus, the range of compositions inferred from astronomical observations for rocky exoplanets implies enormous differences in their dynamics and evolution, due to (i) differences in viscosity and other physical properties, (ii) differences in the composition and buoyancy of crust, (iii) differences in internal heating, and (iv) differences in oxygen fugacity.

### 3.5 Remaining limitations

The validity of the points presented in this work is subject to the principal assumptions that the chemical evolution of the galaxy can be quantitatively modelled, and that the devolatilization technique of Wang et al. (224) can be applied to planets around stars other than our Sun. "Sun-like" stars with vastly different metallicities from our star may have planets that undergo different the condensation/evaporation processes. The accuracy of the approach to quantifying exoplanet compositions is also subject to the dynamically formation and thermal history of such planets, as well as the starting complement of of short-lived radionuclides (e.g. $^{26}Al$) in the system. Without any other information, Earth values must be assumed. An active area of research at the present time is to test for correlations between host star metallicity and the occurrence rate of small/rocky planets (239-243). Most Main Sequence stars are binaries, so the extension of analysis to the effect of stellar multiplicity on the occurrence rate of planets (244-246) is also on-going.

## 4. Conclusions and the new field of Geoastronomy

This work offers path to answering the question of what makes a silicate+metal (terrestrial-type) planet suitable for the *Lavosier-Lomonosov Bridge* and thus, the origin of life. In doing so, we are reminded that the fields of geology and chemistry were combined in the late 19$^{th}$ and early 20$^{th}$ centuries CE to create a new discipline – *Geochemistry* – that gave us such discoveries as the age of the Earth. At about the same time, geology and physics teamed up to become *Geophysics*, which provided us with, among other things, the structure of the Earth's interior and plate tectonics theory. Geology and biology have been combined into *Geobiology*, which explores the intersection of the geosphere with the biosphere. It is now time to consider the merger of Geology and astronomy into a new field – *Geoastronomy* – that seeks to make sense of how planets function within the backdrop of the physical and chemical evolution of the cosmos. Life is, after all, a product of this cosmic evolution. To help in our understanding of the biological phenomenon, we recall the Fundamental Dogma of prebiotic chemistry, that:

- (i) Natural cosmo-geochemical agents reacted together in a planetary environment to make increasingly complex chemical systems (Chemical evolution)
- (ii) Chemical evolution preceded Darwinian evolution
- (iii) Darwinian informational molecules were encapsulated to make early cells
- (iv) Last Common Universal Ancestors of all life came later, and
- (v) This happened when the planets (Earth, Mars? Venus?) were young
- (vi) Conditions for the origin of life are present wherever liquid water and energy are available

Approximately 10% of Main Sequence stars in the galaxy are "Sun-like" (FGK). Statistical analysis suggests that nearly all Sun-like stars host planets (247). Of these, approximately 1-in-3 (or ~$6 \times 10^8$ planets), may host conditions that are considered "habitable" (8). There are a lot of planets to where life could have taken hold just in our Galaxy.

To host a biosphere, an Earth-like exoplanet ought to have:

- (i) a mass at least that of Mars (0.1 $M_\oplus$) to have sufficient internal heat to regenerate crust and retain an atmosphere and hydrosphere in the face of stellar luminosity evolution, but not be too large (≤6 $M_\oplus$) such that it flips to becoming a sub-Neptune by effective capture and retention of disk gas;
- (ii) a stable orbit such that it can survive for at least many millions or better, billions of years;
- (iii) a credible dynamical path of origin; and,

(iv) a cosmochemically tenable composition comprised of starting components that are available at appropriate nuclide abundances in the ISM.

Over billions of years, the galaxy has chemically evolved so that the products of stellar nucleosynthesis are mixed-in and go into forming new star-planet systems with slightly different initial conditions. Planets forming around other stars now are different that those that emerged in the early galaxy. Our Solar System appeared after the Milky Way has already matured, more than 9 billion years after the Big Bang. Hundreds of millions of terrestrial-type planets exist in our galaxy which orbit stars that can be broadly classified as "Sun-like" (FGK). The compositions of these stars derived from devolatilization analysis of spectra and verified by GCE models can be used to place bounds on the compositions of the rocky planets that orbit them.

Although no experiments have thus far succeeded in crossing the *Lavosier-Lomonosov Bridge* from the "inanimate" to the "animate" world, and even if they succeed one day in doing so does not mean that is actually how it happened on Earth or elsewhere, at least we can say that the reactants and conditions necessary for this process to occur on cosmochemically Earth-like exoplanets are abundant. Life, at least microbial life, on or in worlds around other stars, and even in rogue objects drifting in interstellar space, should probably be viewed as a certainty.

………..

**(13370 words)**

**(254 references)**

**(22 Figures)**

**(1 Table)**

**ACKNOWLEDGEMENTS**

I extend sincere gratitude to colleagues Elizabeth Frank, Brad Meyer, Haiyang Wang, Maria Lugaro, Kevin Heng, Jens Hoeijmakers, Paul Tackley, Maxim Ballmer, Robert Spaargaren, Georgios Perdikakis, Ramon Brasser, Marc Hirschmann, Gregor Golabek, Tim Lichtenberg, David Rubey, Daniel Frost, Craig O'Neill, Oleg Abramov, Steven Benner, Thomas Carrell, Dieter Braun and Mark Harrison for useful discussions revolving around the general concept of Geoastronomy and topics intimately relevant to the *Lavosier-Lamontsov Bridge* in particular. I am deeply grateful to my colleagues Christian Koeberl, Manuel Guedel, Joao Alvez and the University of Vienna, Department of Lithospheric Research (Vienna, Austria) for the Ida Pfeiffer Guest Professorship at which time this manuscript was completed. Thanks also go out to the Alexander von Humboldt Foundation for the Humboldt Research Prize which hosts me for part of the year at the Friedrich Schiller University (Jena, Germany) to work with Christoph Heubeck. The Research Centre for Astronomy and Earth Sciences (Budapest, Hungary) funds the Origins Research Institute (ORI).

**Figure 1**: Habitable volumes of different Solar System objects (blue) defined by the stability of liquid water at their surfaces or interiors, compared to equilibrium temperatures (K; red) as a function of heliocentric distance (astronomical units, AU). Earth's total habitable volume ($10^{10}$ $km^3$) is subordinate to cumulative value ($10^{12}$ $km^3$) of all large (>800 km diameter) Trans-Neptunian Objects (TNOs) that exist beyond Pluto at 40 AU.

**Figure 2**: Pressure (Pascals, Pa; $10^{-5}$ bar = 1Pa) vs. Temperature (K) phase diagram for pure water. The thermodynamic properties of water yield insight on the availability for biochemistry at the surfaces of the terrestrial planets. Shown are values that indicate Mars's surface (red box) is near the triple point but its subsurface overlaps considerably with the terrestrial habitable field, common Earth life (green box) is well-within the liquid field but the allowable space for microorganisms (yellow box) is very large. Venus' atmosphere is bound by its surface near the critical point and its upper atmosphere (at 48-65 km altitude) shows volumes of overlap with that of Earth and Mars. Ice polymorphs: (h) = hexagonal, (o) = orthorhombic, (c) = cubic ice structures, respectively.

**Figure 3:** Periodic table of elements essential to life. Bulk elements (shaded orange) are structural components of cells and require both abundance and bio-availability in the environment for function. Trace elements (shaded bright yellow) like iron, cobalt, magnesium and nickel are key to biological processes, whereas Cu and Zn are less so. Although the trace elements represent a very small fraction of the mass of a cell, all are essential to life mostly because of their central catalytic roles in specific proteins, including enzymes.

**Figure 4:** Stellar evolution of low-mass (left cycle) and high-mass (right cycle) stars, with examples of observed or inferred objects labelled in italics. Credit: NASA Goddard Space Flight Center.

**Figure 5:** Computed mole fraction of volcanic gases at 1 bar (100 kPa) and 1800 K with a mass H/C ratio of 0.5 (modified from ref. 44) for planets of broadly Earth-like mantle composition. Solid lines show the calculated values at 1800 K as a function of oxygen fugacity ($f\,O_2$) expressed in $\log_{10}$ units vs. co-existing iron and wüstite ($\Delta$ IW) from the definition of O'Neill (1988). The IW buffer (vertical dotted line) is approximately 3.5 $\log_{10}$ units below QFM. The oxygen fugacity values apply only to the 1800 K calculation, whereas the 500 K values are via simple cooling from 1800 K gas. The shaded (cyan) region corresponds to the range of Earthl samples (Kilauea and MORB glasses, Abyssal peridotites, continental xenoliths) of mantle melts and upper mantle compositions. More reducing mantles in the Solar System, such as those of the Moon and Mars (from lunar samples and Mars/SNC meteorites, respectively), may be analogues to terrestrial-type exoplanets of masses <0.8 $M_{\oplus}$ (grey shaded region).

**Figure 6:** Postulated stages in the origin of life (ref. 83). Prebiotic chemistry produced the building blocks of RNA in the natural environment(s) of the young planet the was formed from the nuclides accumulated in Galactic Chemical Evolution (GCE). After cooling, prebiotic chemistry could commence. The first stage of proto-biological evolution in this scenario begins with RNA molecules that can perform catalytic functions needed to assemble themselves from available building-blocks in their immediate environment. These RNA molecules then evolve as self-replicators that recombine and mutate, and adapt, in the face of selection pressures to occupy new environmental niches. Subsequently, these RNAs develop an increasingly complex suite of enzymatic activities. The next stage may have been when RNAs incorporated wild proteins synthesized abiotically, ultimately synthesizing their own and began the process of RNAs replicating with other RNAs. Ultimately, DNA appeared as the stable information center with error-correction, but still with mutation and recombination. The crystallization of the genome of life occurred at the time of the universal cenancestor cells (a.k.a. Last Universal Common Ancestors; LUCA); a community of organisms that are rooted in the Bacteria domain. (Phylogenetic Tree graphic generated from https://itol.embl.de/).

**Figure 7:** Discovery date vs. estimated masses for planets orbiting Main Sequence stars, with $\log_{10}$ y-axis of estimated masses and year of discovery running on the x-axis. The lower envelope of this semi-log plot shows a "Moore's Law"-like progression on how planets with masses close to that of Earth are being discovered. Transiting planets are shown in red, planets detected via radial velocity are in blue. Shown in the cyan field is the range of masses attributable to "terrestrial-like" planets in the mass range 0.1 – 6 $M_{\oplus}$. Data from: https://exoplanetarchive.ipac.caltech.edu.

**Figure 8: (A)** Evolution of planetary mantle concentrations (ng $g^{-1}$) of the long-lived heat-producing radionuclides $^{235}$U, $^{238}$U, $^{232}$Th and $^{40}$K, for Earth at time since Solar System formation. **(B)** Change in mantle heat production (expressed in $10^{-12}$ W $kg^{-1}$) based on starting mantle concentrations in **(A)** of Earth history run over $10^{10}$ y. Note the short term demise of $^{235}$U, followed by $^{40}$K, the intermediate role of $^{238}$U in the first ~5 Gry. The terminal long-term overtake of $^{232}$Th represents the principle radiogenic heat source for cosmochemically Earth-like planets older than ~5 Gyr (modified from ref. 141).

**Figure 9:** Solar photosphere – CI vs. A (proton number + neutron number) for the nuclides. It can be seen that over forty elements have photospheric - CI chondrite abundance within a few percent of one another (refs. 154-156). It is also worth mentioning that the relative elemental abundances of Fe, Mg, and Si are similar among the Sun, Earth, Mars, the Moon and meteorites (see ref. 248).

**Figure 10:** Logarithmic abundances of present-day nuclides in the Sun plotted as mass fraction against mass number (A). The even-odd mass numbers and isotopes of the nuclides are plotted. The color coding of the nuclide groups corresponds to different possible nucleosynthetic paths shown in the legend. Capitalization denotes the major nucleosynthetic process, whereas lower-case corresponds to a minor nucleosyntetic pathway. U = cosmological/Big Bang synthesis; H = Hydrogen burning; X = cosmic ray spallation; He = Helium burning; N = hot/explosive hydrogen burning; C = Carbon burning; Ex = explosive nucleosynthesis; Ne = Neon burning; O = Oxygen burning; S = s-process; E = nuclear statistical equilibrium; R = r-process; P = p-process; RA = r-process producing actinides.

**Figure 11:** Schematic illustration showing the flow of mass through the galactic disk and the pathways of matter from gas+dust to star and back to ISM. Important structural features of the galactic environment discussed in the text are labelled. SMBH = super-massive black hole at the galactic center (ref. 249).

**Figure 12:** Computed ISM gas mass fractions of the major mantle-forming elements (Fe and K are treated separately) over time. In the analytical GCE model, these nuclides are constantly enriched over galactic history in the disk.

**Figure 13:** Co-evolution of [Mg/Si] vs. [Fe/H] normalized to solar values.

**Figure 14:** Metallicity expressed in [Fe/H] vs. [Mg/Si] (dex units) for known planetary host stars (cyan) and stars with no documented giant planets (filled); data from ref. 212 and ref. 250). Solar values are shown (156). Note the trend for giant planets around high metallicity stars (cf. ref. 251) also shows a tendency of values toward [Mg/Si] > solar.

**Figure 15:** Model of the mass fraction evolution of $^{56}$Fe compared to,Mg and Si in the ISM disk. Initially, Fe is only produced in massive stars, but about 1 Gyr into evolution, Type 1a supernovae begin contributing iron to the gas, boosting its production rate compared to other nuclides.

**Figure 16:** Preliminary model of the mass fraction evolution of $^{40}$K compared to total K in the ISM.

**Figure 17:** Observed variations in stellar compositions of FGK stars from spectrographic measurements (translucent symbols) plotted in a ternary of MgO, $SiO_2$, FeO. Elemental compositions are from the Hypatia catalogue (252) and transmuted to oxides. Solar-system composition is marked by the yellow diamond (156) and Earth by the green dot (191). Proposed fields for broadly pyroxene (px)-dominated vs. olivine (ol)-dominated mantle mineralogies in rocky planets expected to exists around these stars given these compositional parameters are also shown. Slopes of the Mg/Si and Fe/Si ratios are indicated.

**Figure 18:** Predicted mass fractions of the long-lived radionuclides responsible for heat production in the interiors of rocky planets in the galaxy. This model output assumes constant enrichment and accounts for the mixing time, or approximately the half-rotation, of the galaxy (~125 Myr). The model also accounts for instantaneous decay of each species, but secular equilibrium is reached owing to the relatively long half-lives (>700 Myr) compared to the galactic period. The nuclide 235U is already at equilibrium concentration.

**Figure 19:** The anticorrelation of FeO and MgO in planetary mantles as the function of core size (by mass, at zero pressure). This model is only for volatile-free bulk CI-composition planets. (157). Earth values are labelled. ***Inset***: Cross-section of the Earth showing its layered structure. CMB, core–mantle boundary; ICB, inner-core boundary; Mass %, percentage of mass counting from the center; Volume %, percentage of volume counting from the center. (based on ref. 253).

**Figure 20:** The volatility trend, illustrated by the wedge in grey, is shown as a comparison of the elemental abundances of the bulk silicate Earth (i.e. the outer part of the Earth, essentially made by silicates) to CI chondritic abundances (normalized to 1) as a function of the 50% condensation temperature (154). The normalization-reference element is Mg. The legend indicates the classification of elements according to their geochemical character: lithophile (elements) preferring to be in the silicate shell/mantle rocks, siderophile tending to be in the metal phase (i.e. core), chalcophile partitioning into sulfides. Above the legend, **HSE** is the abbreviation of highly siderophile elements. **Main comp.** near the upper x-axis denotes Main components, which are elements abundant in the Earth. The figure is from Palme and O'Neill (ref. 157).

**Figure 21:** The elemental composition of solar system rocky bodies, including CI chondrites (157), Earth (223), Mars (207), and Venus (254), normalized to the protosolar elemental abundances in the yellow field, on the equal basis of $10^6$ Al atoms. Taylor reports (207), the elemental composition of Mars as the primitive Martian mantle composition, so only lithophile elemental abundances are plotted here. The abundance differences between terrestrial planets is small to indistinguishable.

**Figure 22:** Using solar values against CI in **Figure 9**, and compared to the 50% condensation temperature relationships plotted in **Figures 20** and **21**, we can then compute the devolatilization trend for the proto-Sun and the Earth. We use this trend to calculate the end-member compositions of terrestrial planets around any star for which spectroscopic data are available, and for which analysis shows terrestrial-type exoplanets exist. To quantify the dual behaviour of the devolatilization pattern shown here, a $\chi^2$ fit of the *f* values to the joint model: $\log(f) = a \log(T_C) + b$ and $\log(f) = 0$ yields best-fit coefficients $a = 3.676 \pm 0.142$ and $b = -11.556 \pm 0.436$. This means that there now exists a quantitative method to determine the bulk compositions of terrestrial-type exoplanets from stellar abundance spectra.

**Figure 1.**

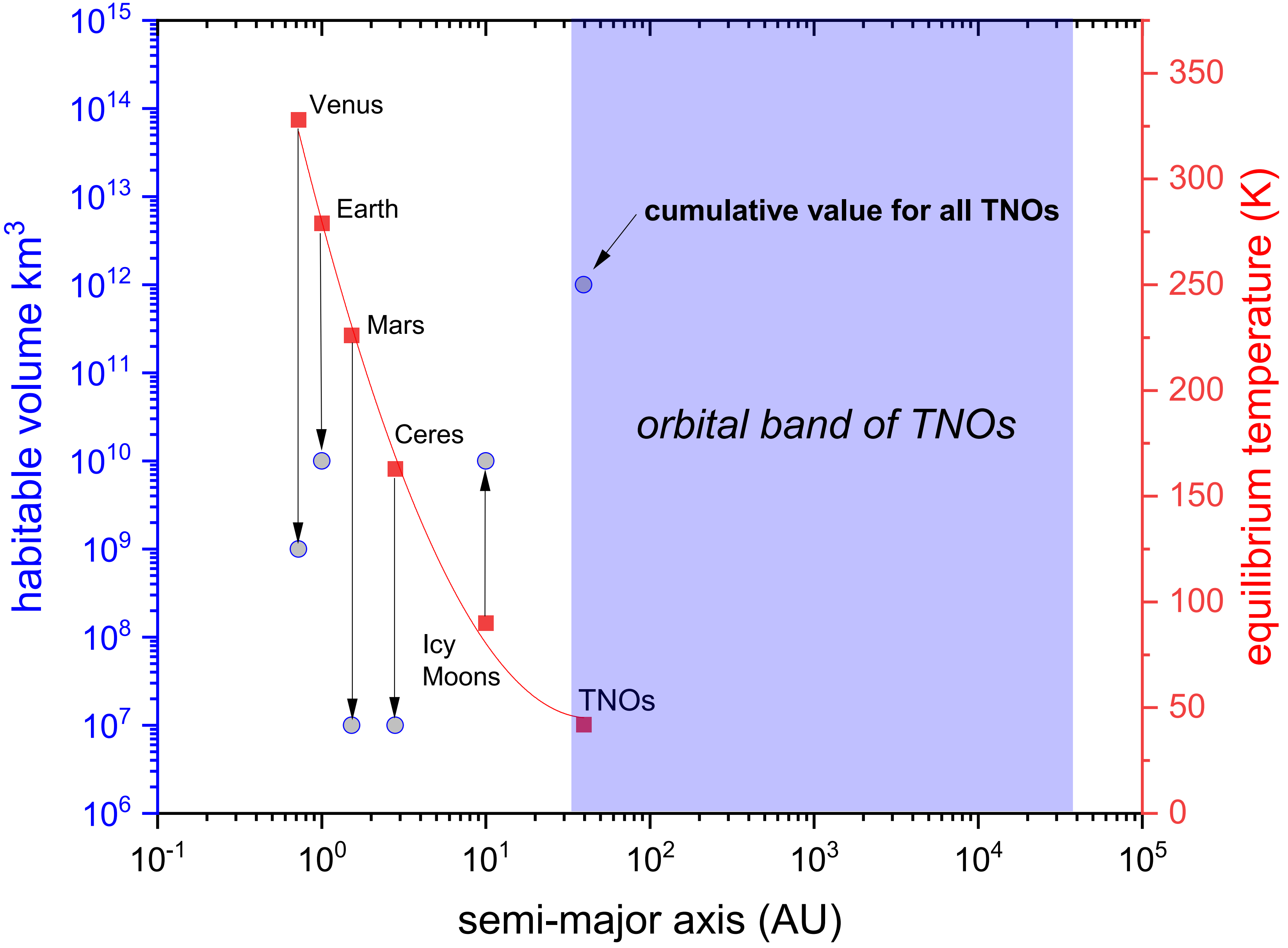

**Figure 2.**

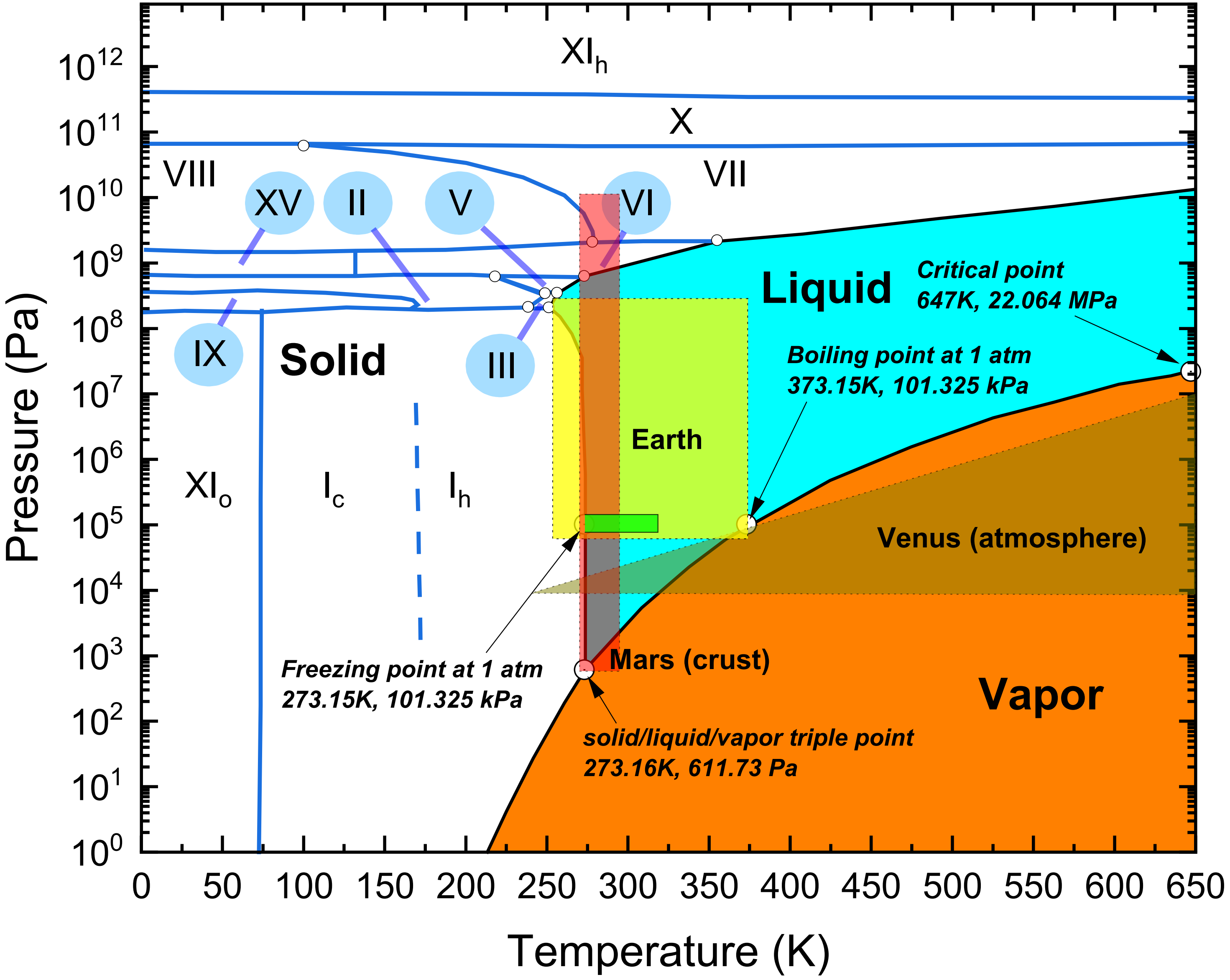

**Figure 3.**

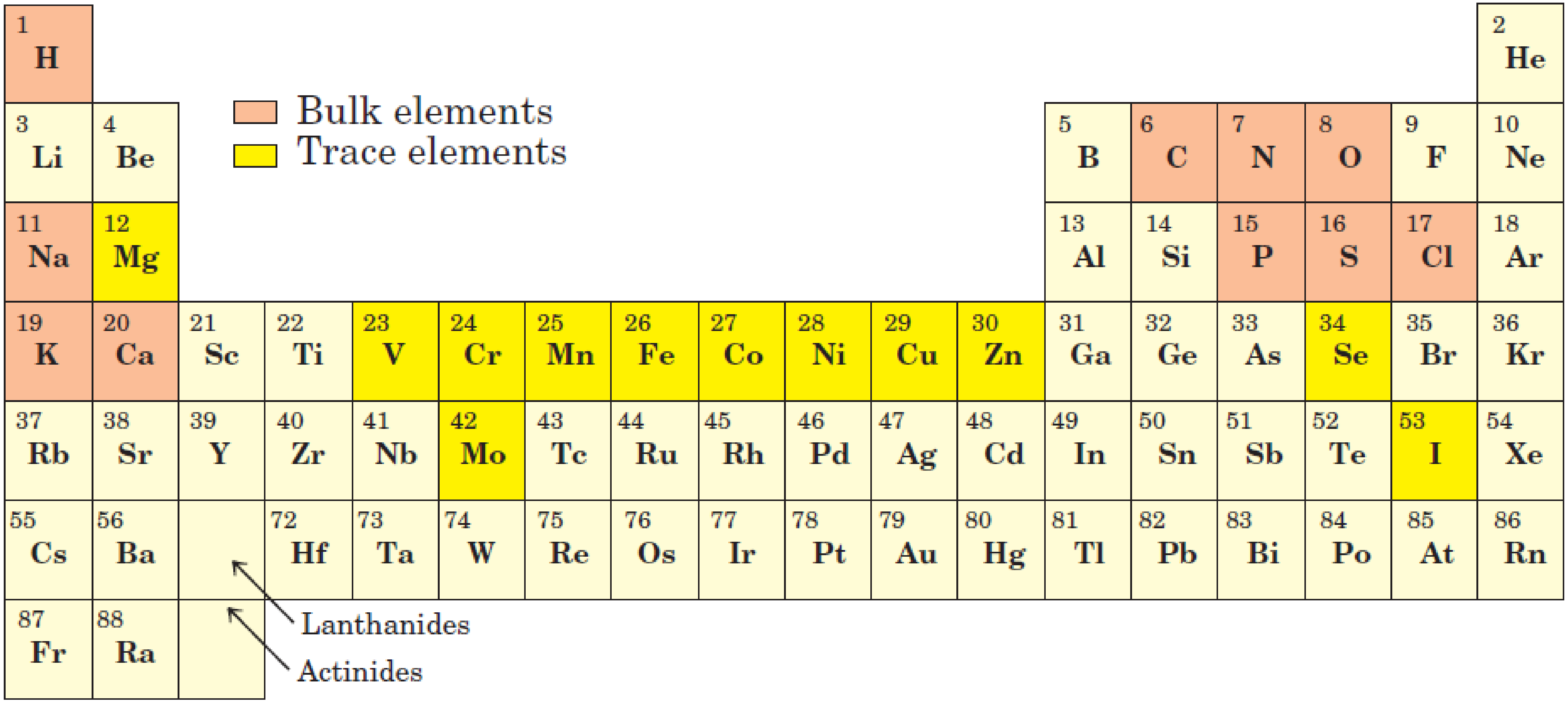

**Figure 4.**

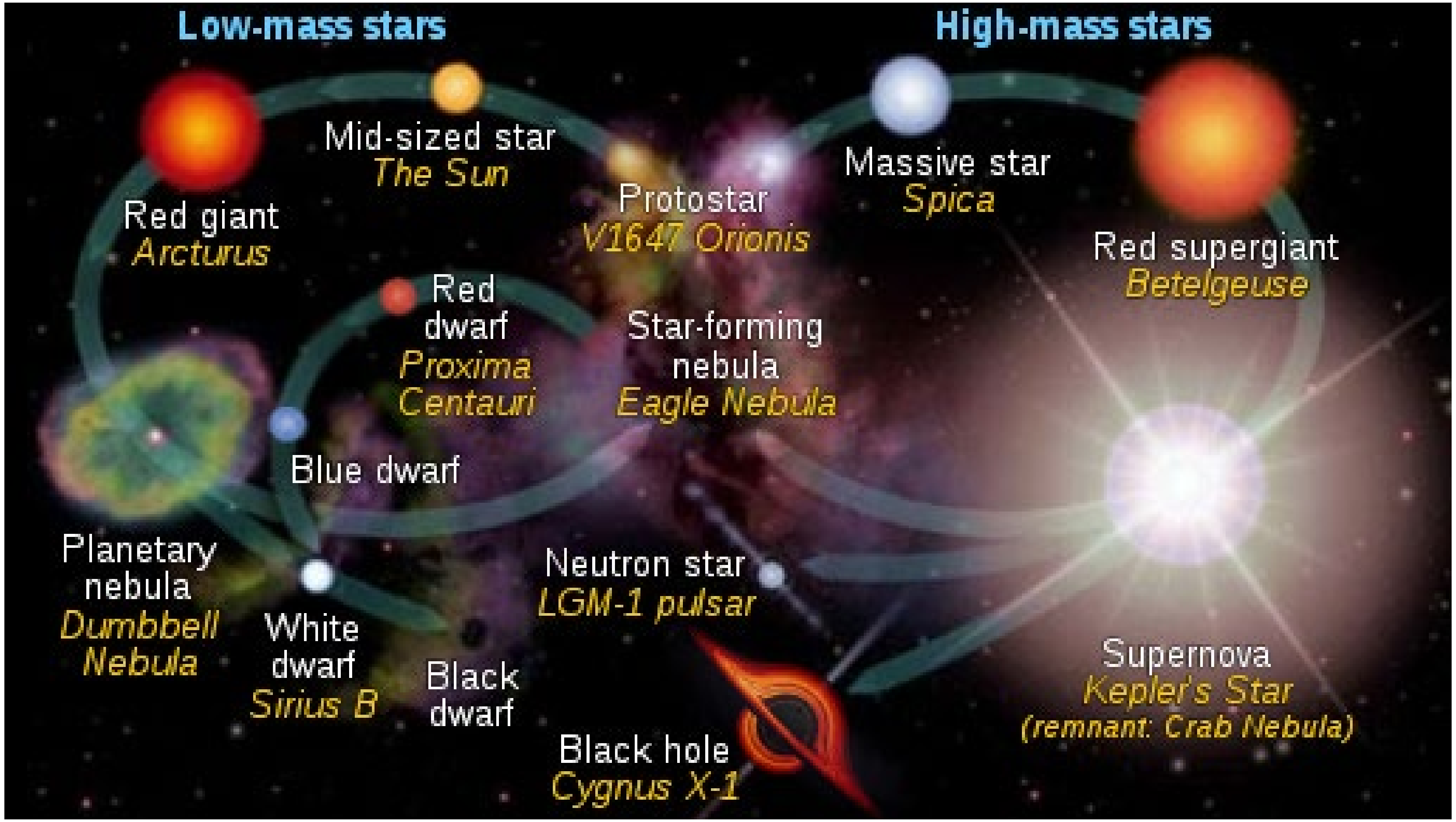

**Figure 5.**

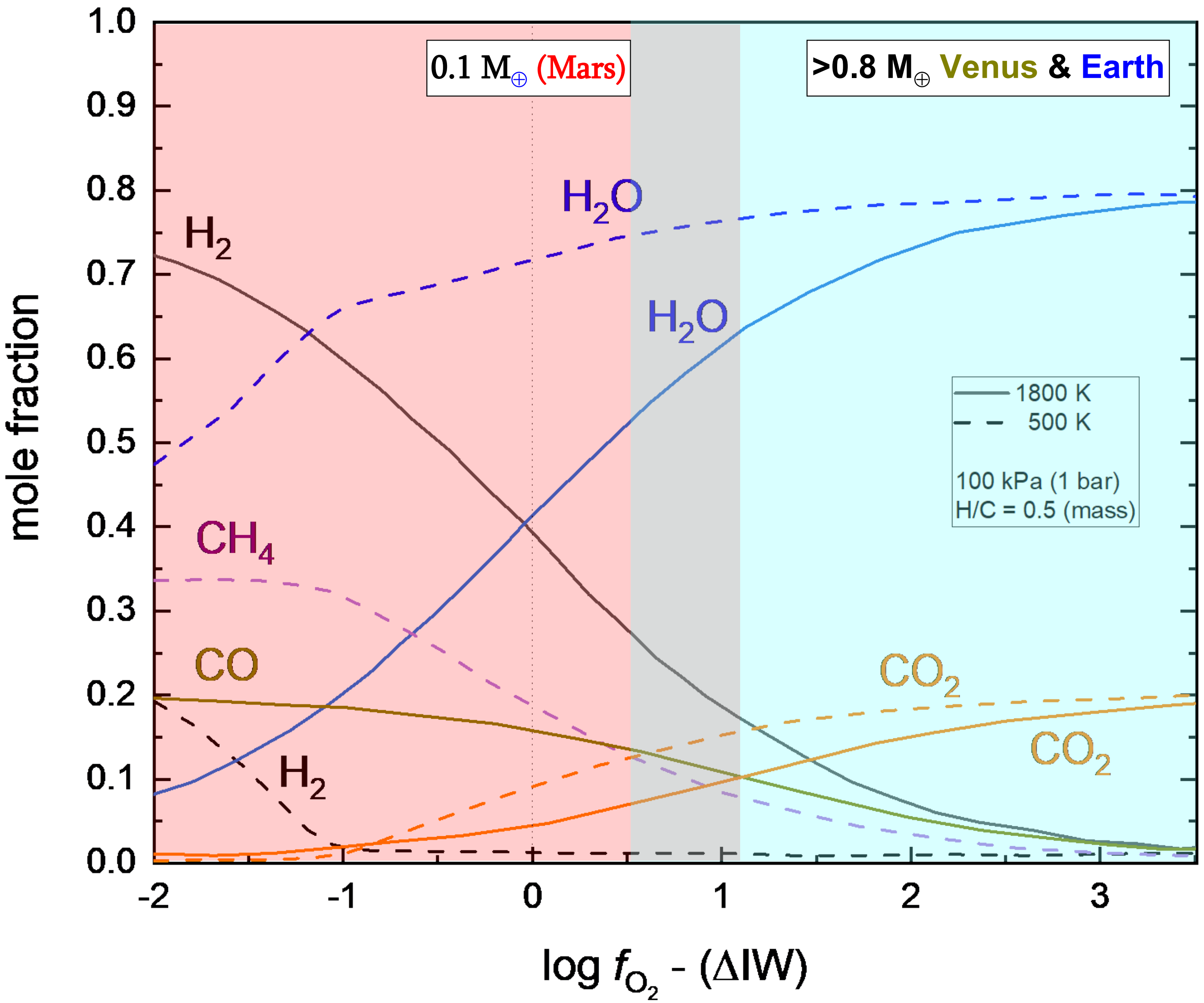

Figure 6.

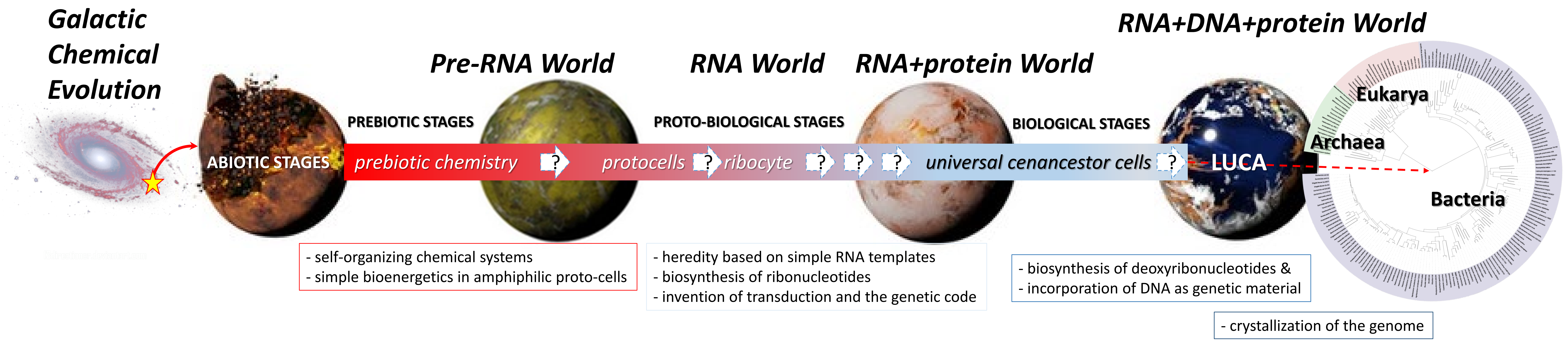

Galactic Chemical Evolution
Pre-RNA World
RNA World
RNA+protein World
RNA+DNA+protein World
PREBIOTIC STAGES
PROTO-BIOLOGICAL STAGES
BIOLOGICAL STAGES
ABIOTIC STAGES
prebiotic chemistry
?
protocells
?
ribocyte
?
?
?
universal cenancestor cells
?
LUCA
Eukarya
Archaea
Bacteria
- self-organizing chemical systems
- simple bioenergetics in amphiphilic proto-cells
- heredity based on simple RNA templates
- biosynthesis of ribonucleotides
- invention of transduction and the genetic code
- biosynthesis of deoxyribonucleotides &
- incorporation of DNA as genetic material
- crystallization of the genome

**Figure 7.**

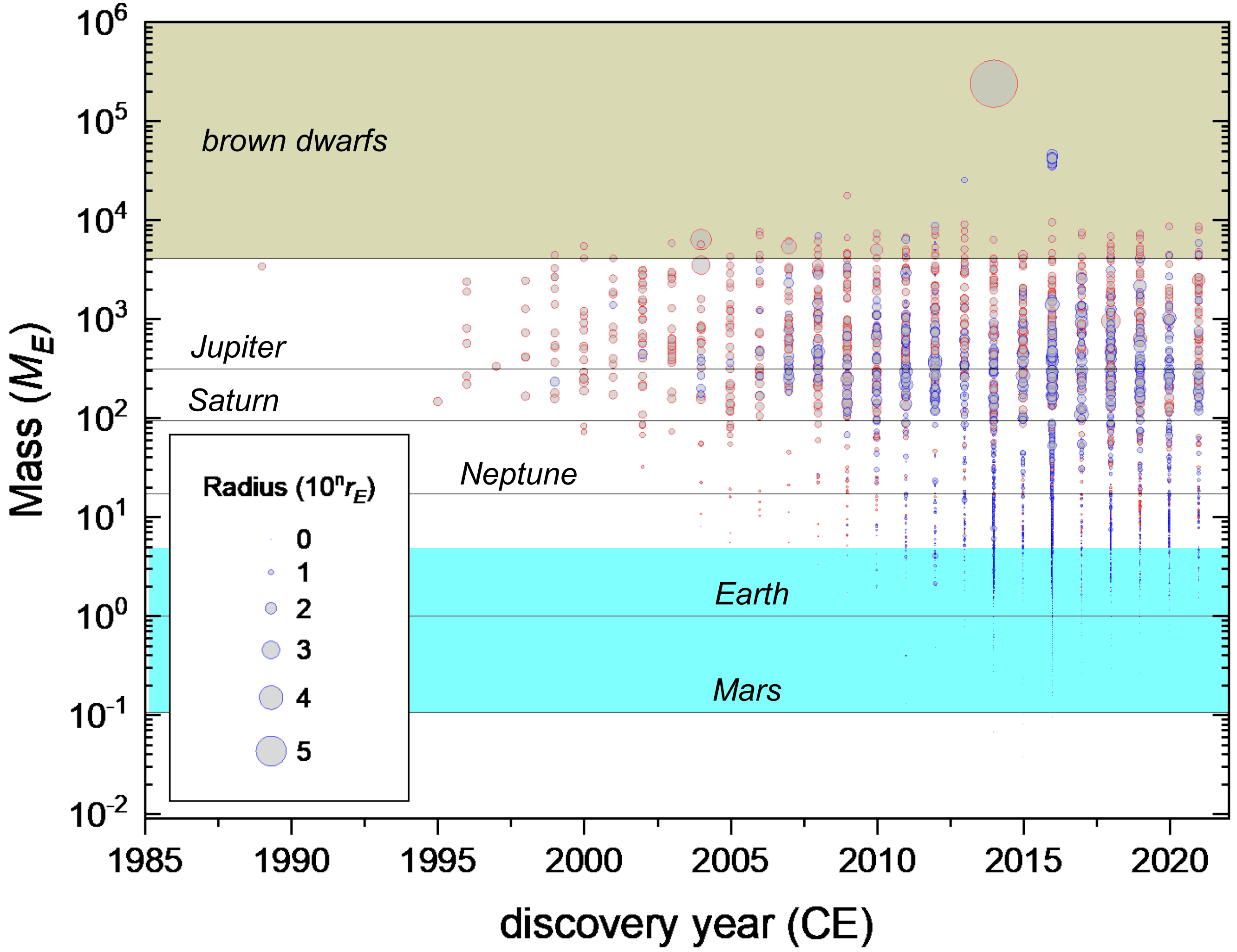

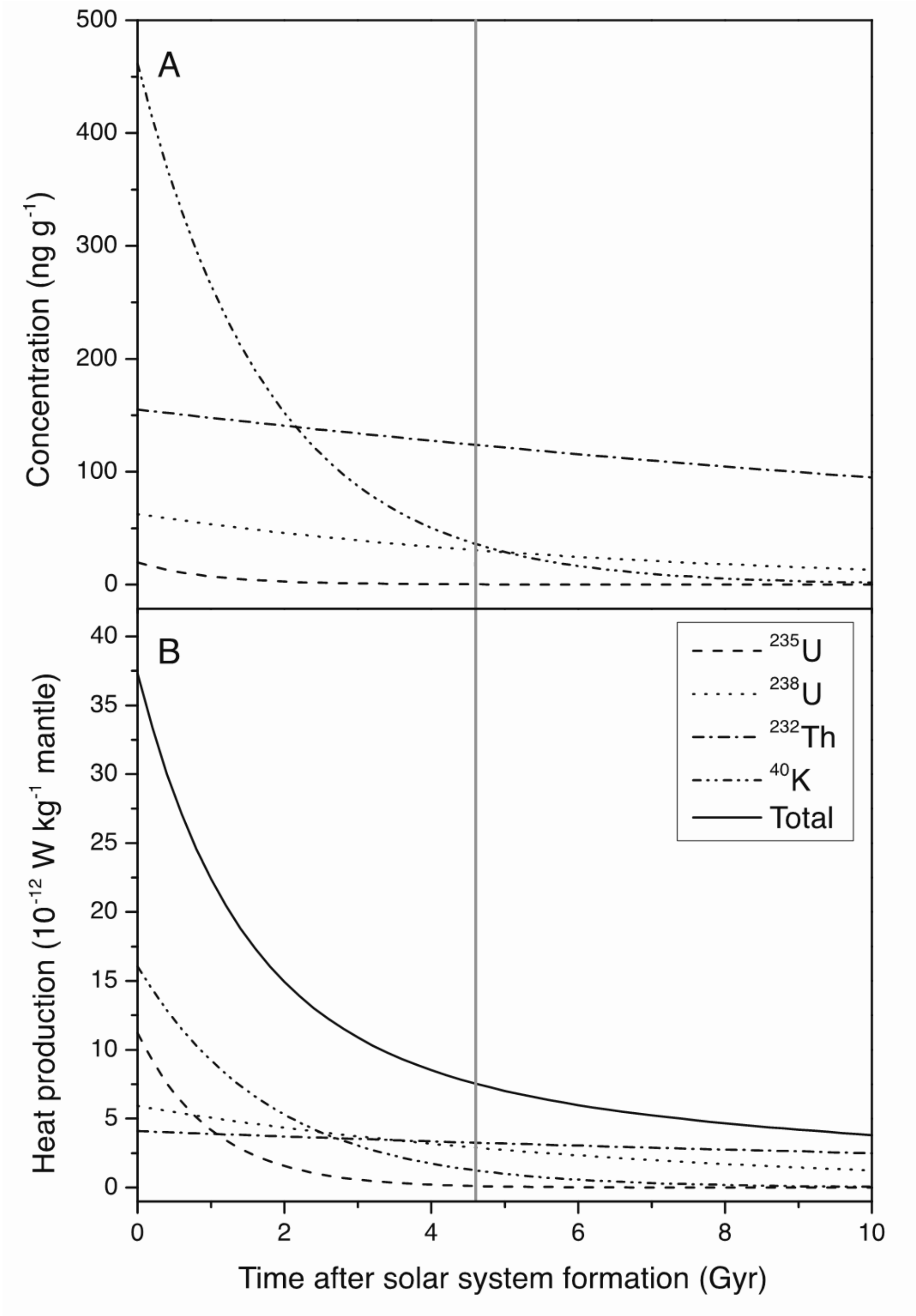

**Figure 8.**

**Figure 9.**

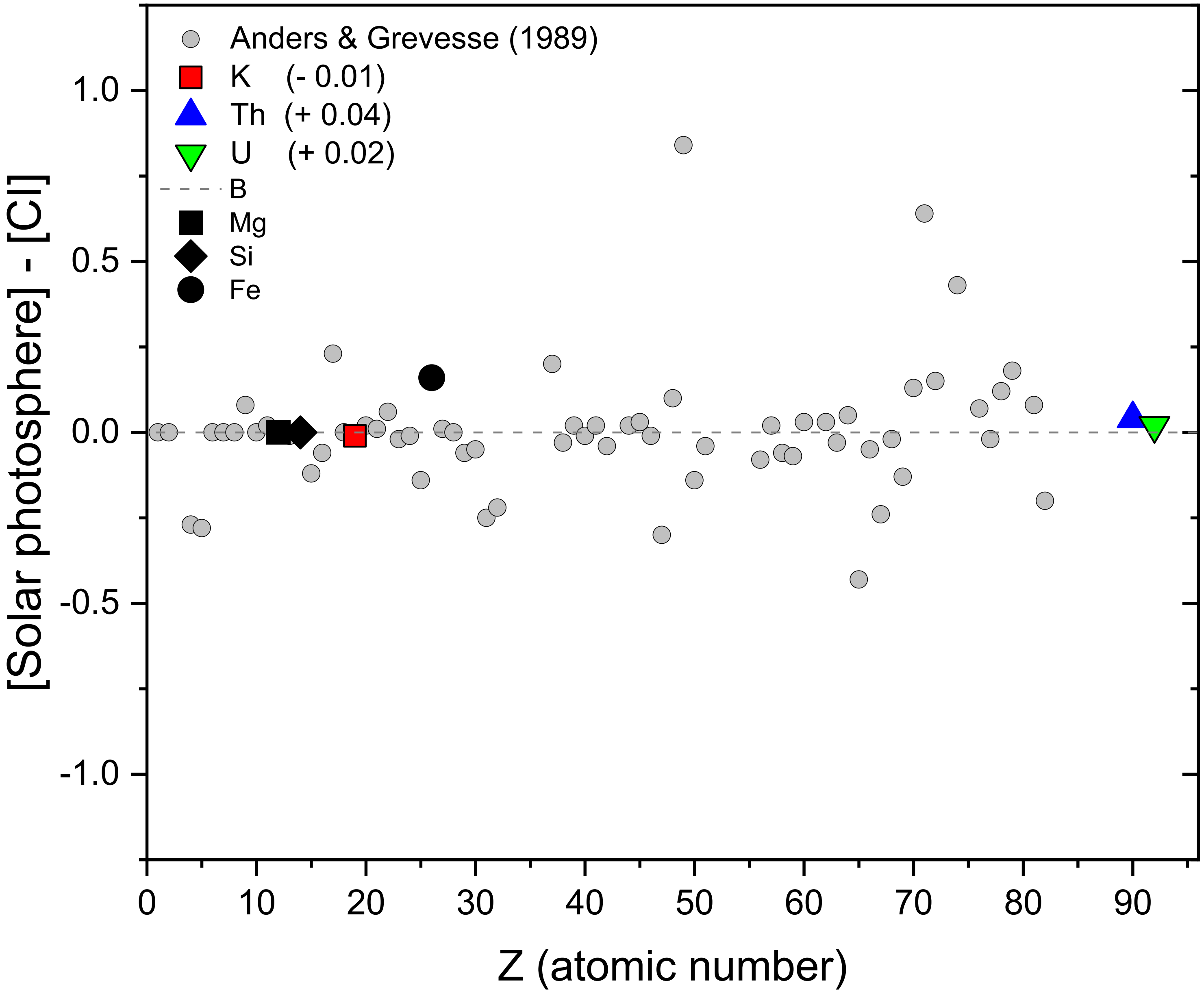

**Figure 10.**

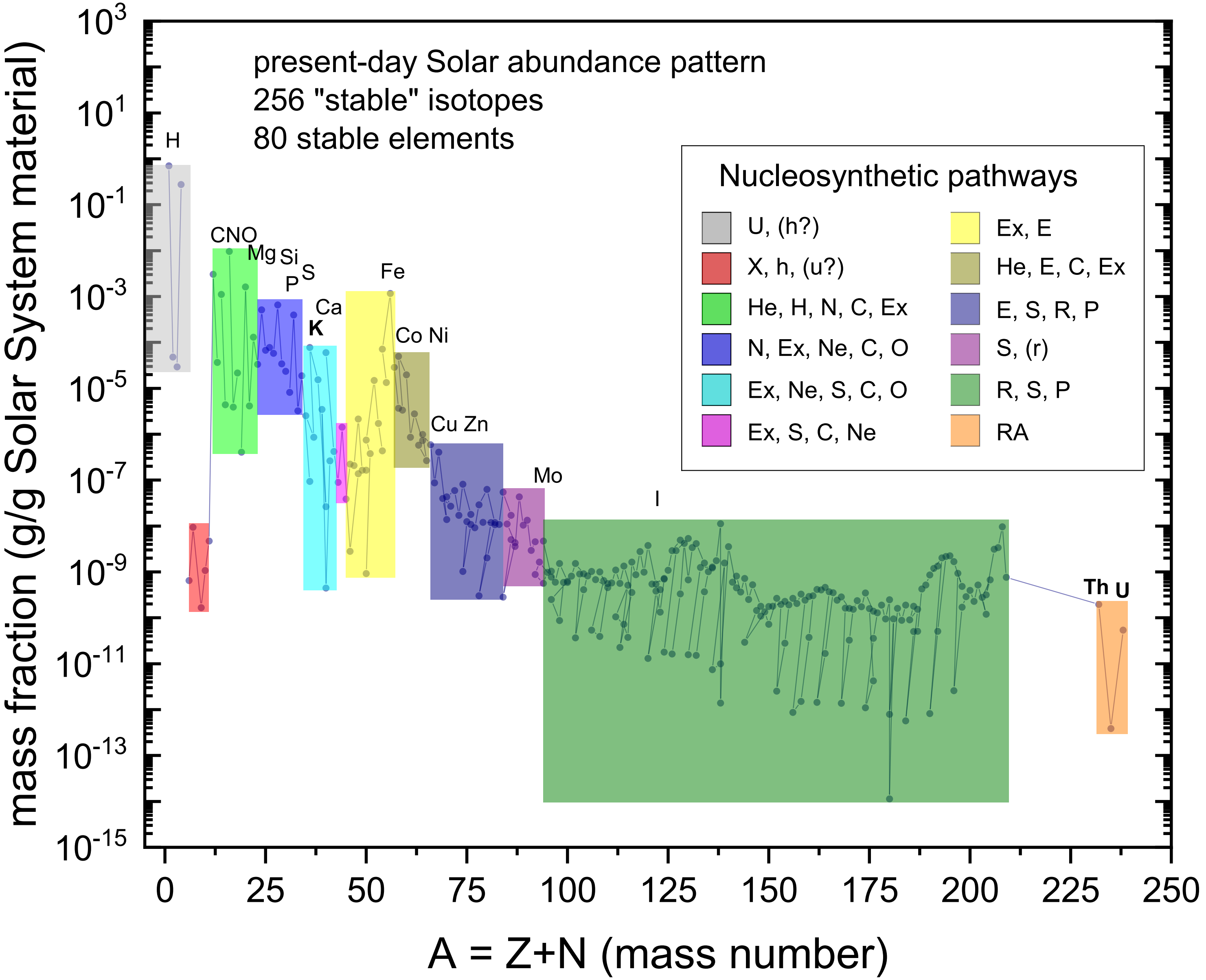
present-day Solar abundance pattern
256 "stable" isotopes
80 stable elements
Nucleosynthetic pathways
U, (h?)
X, h, (u?)
He, H, N, C, Ex
N, Ex, Ne, C, O
Ex, Ne, S, C, O
Ex, S, C, Ne
Ex, E
He, E, C, Ex
E, S, R, P
S, (r)
R, S, P
RA
H
CNO
Mg
Si
P
S
Ca
K
Fe
Co Ni
Cu Zn
Mo
I
Th
U
mass fraction (g/g Solar System material)
A = Z+N (mass number)

**Figure 11.**

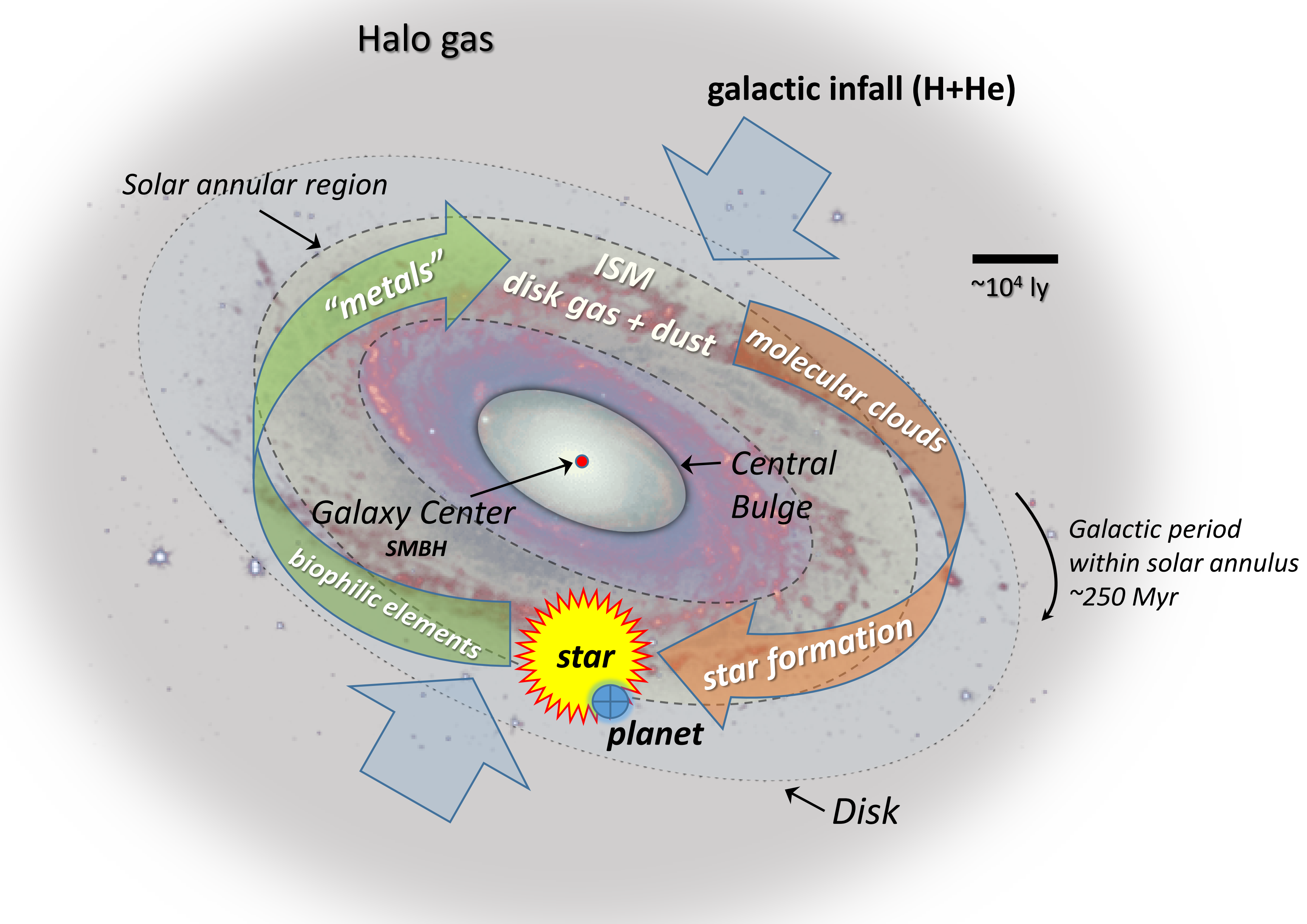

**Figure 12.**

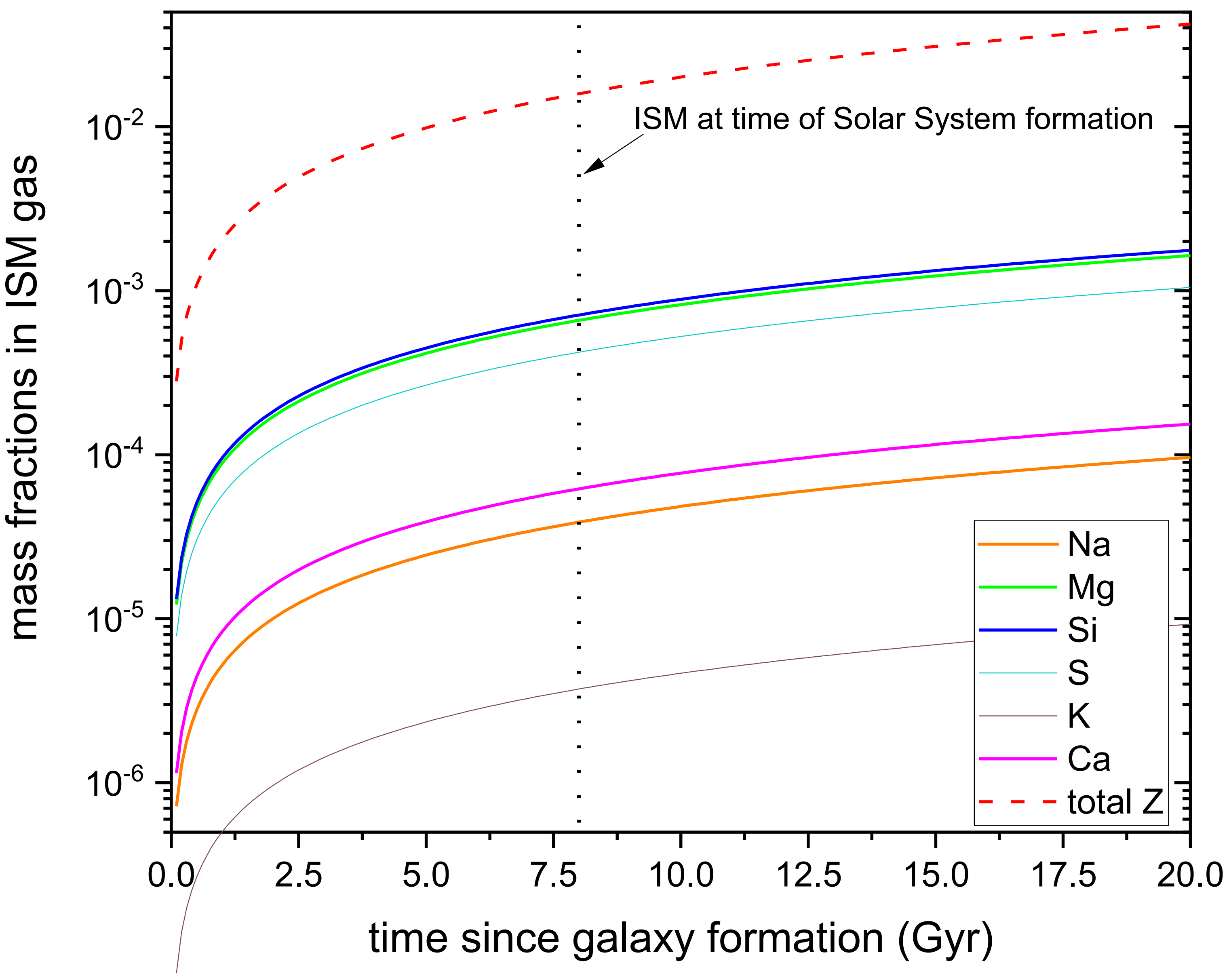

**Figure 13.**

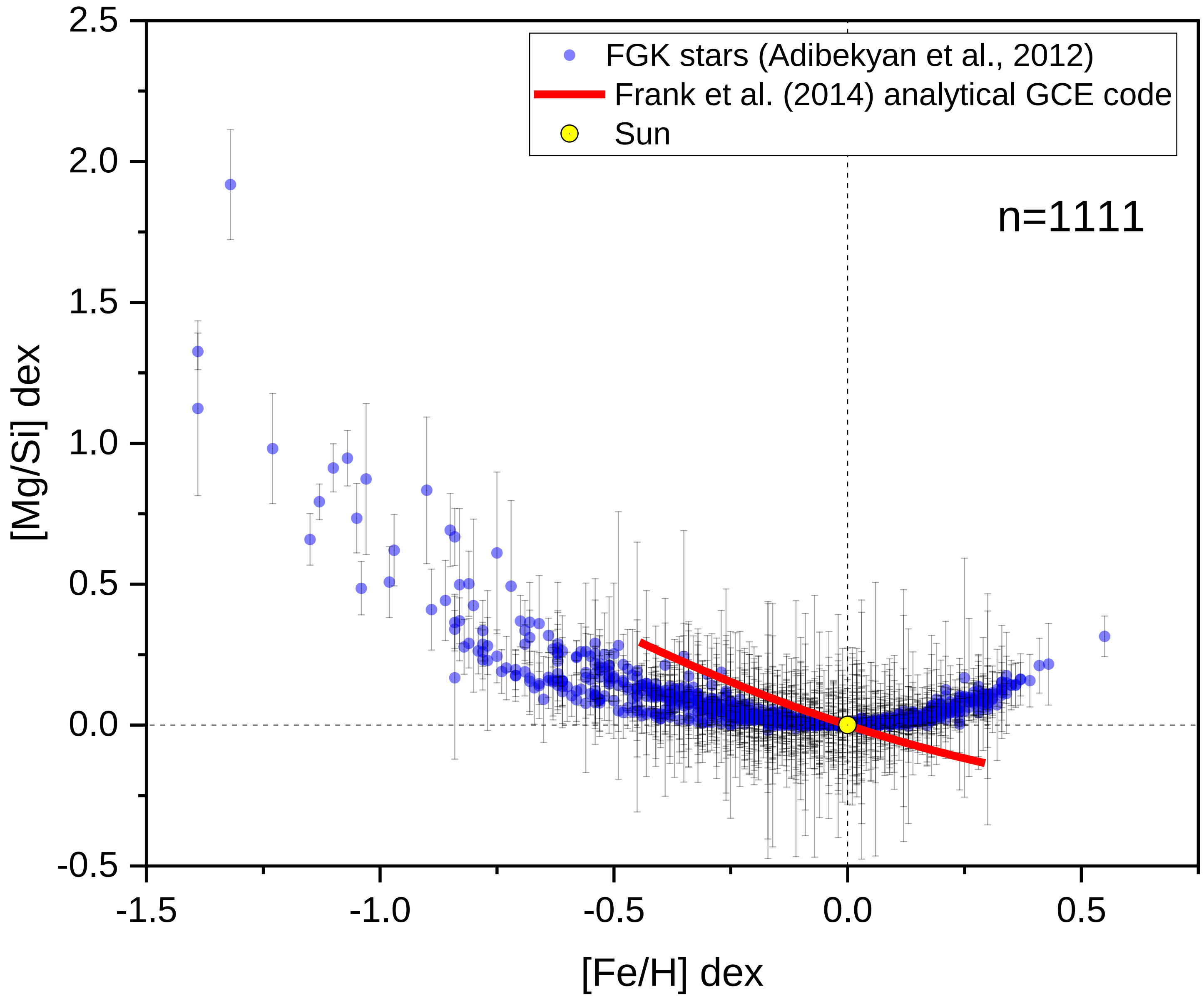

**Figure 14.**

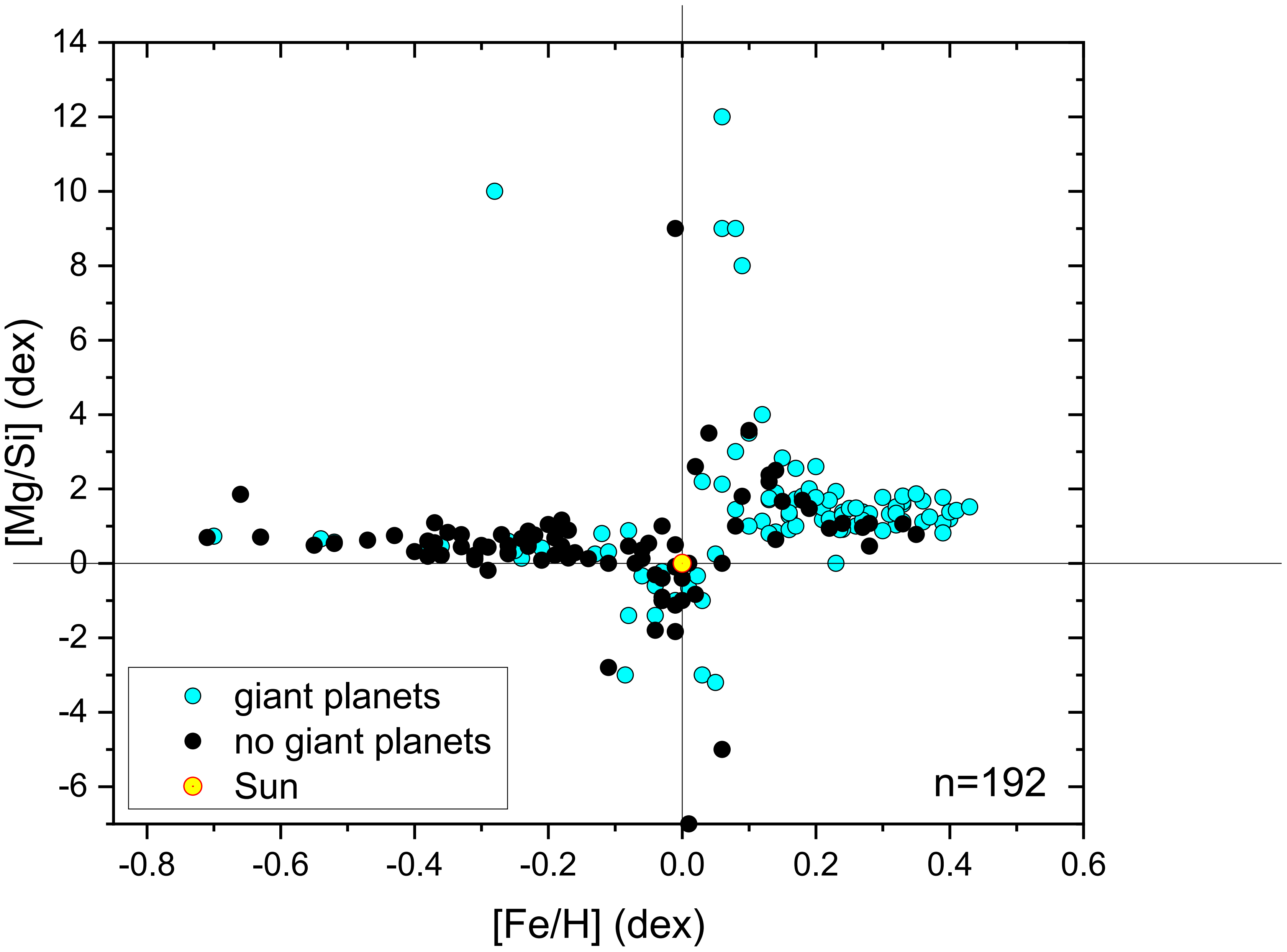

**Figure 15.**

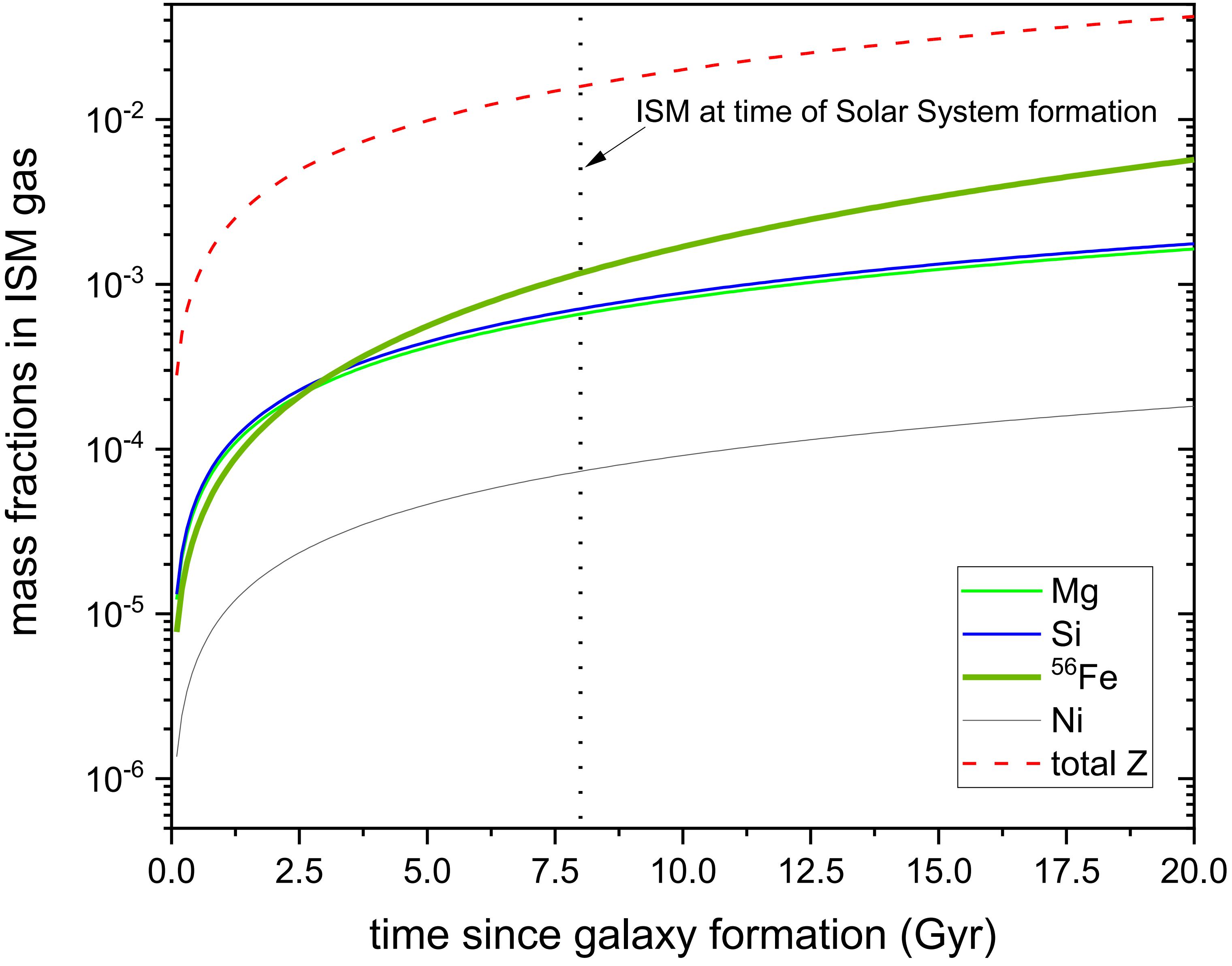

Figure 16.

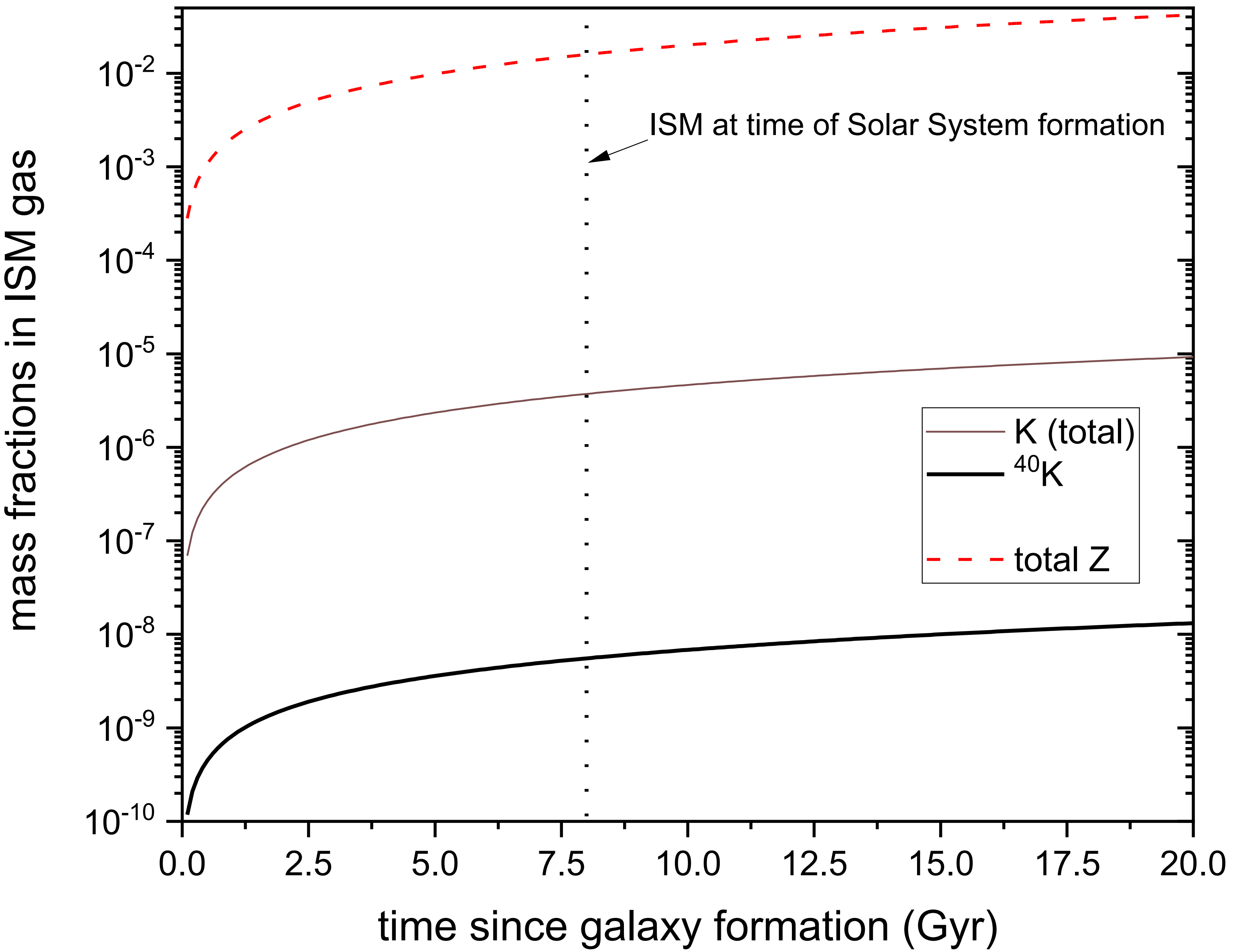

**Figure 17.**

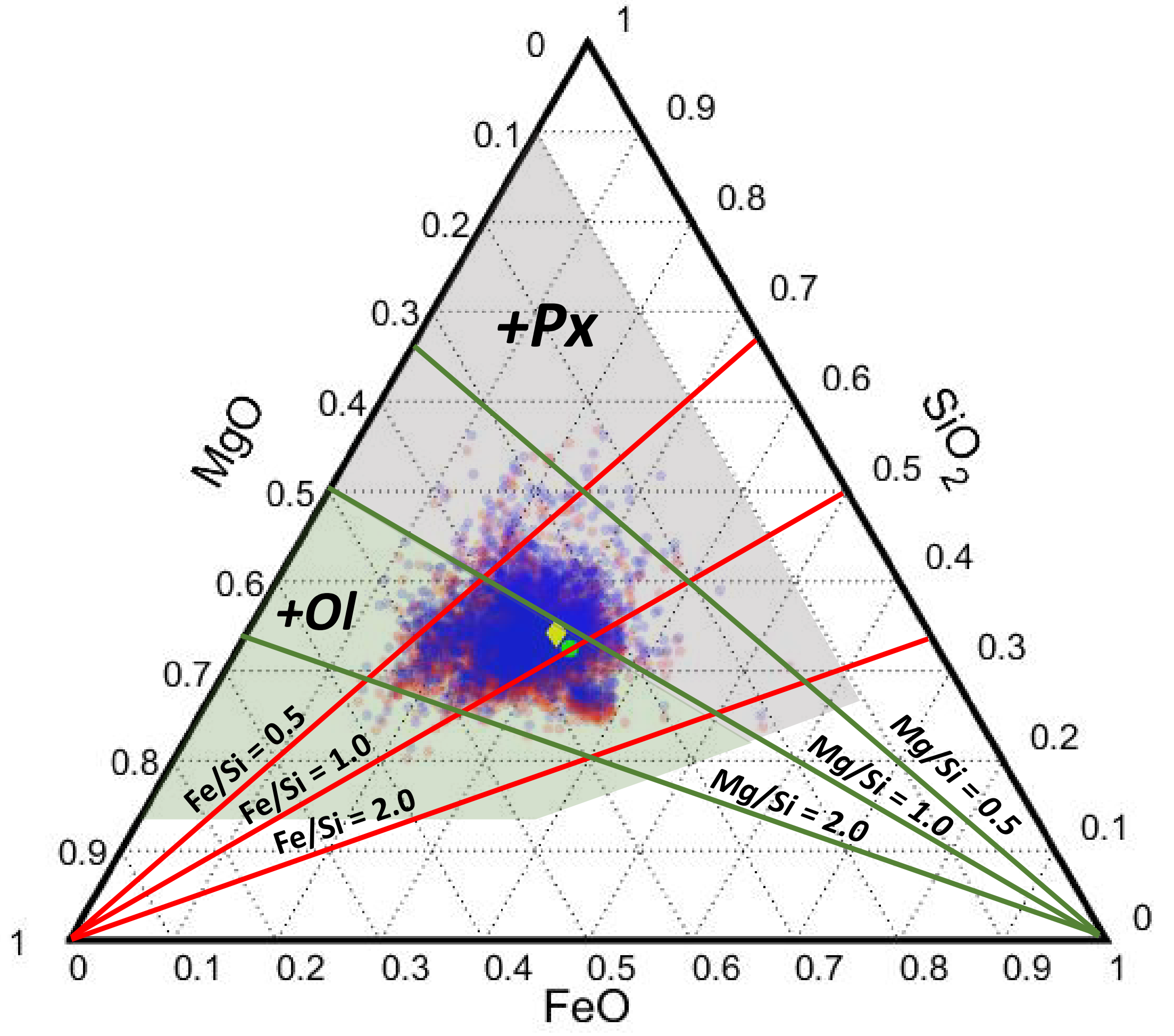

**Figure 18.**

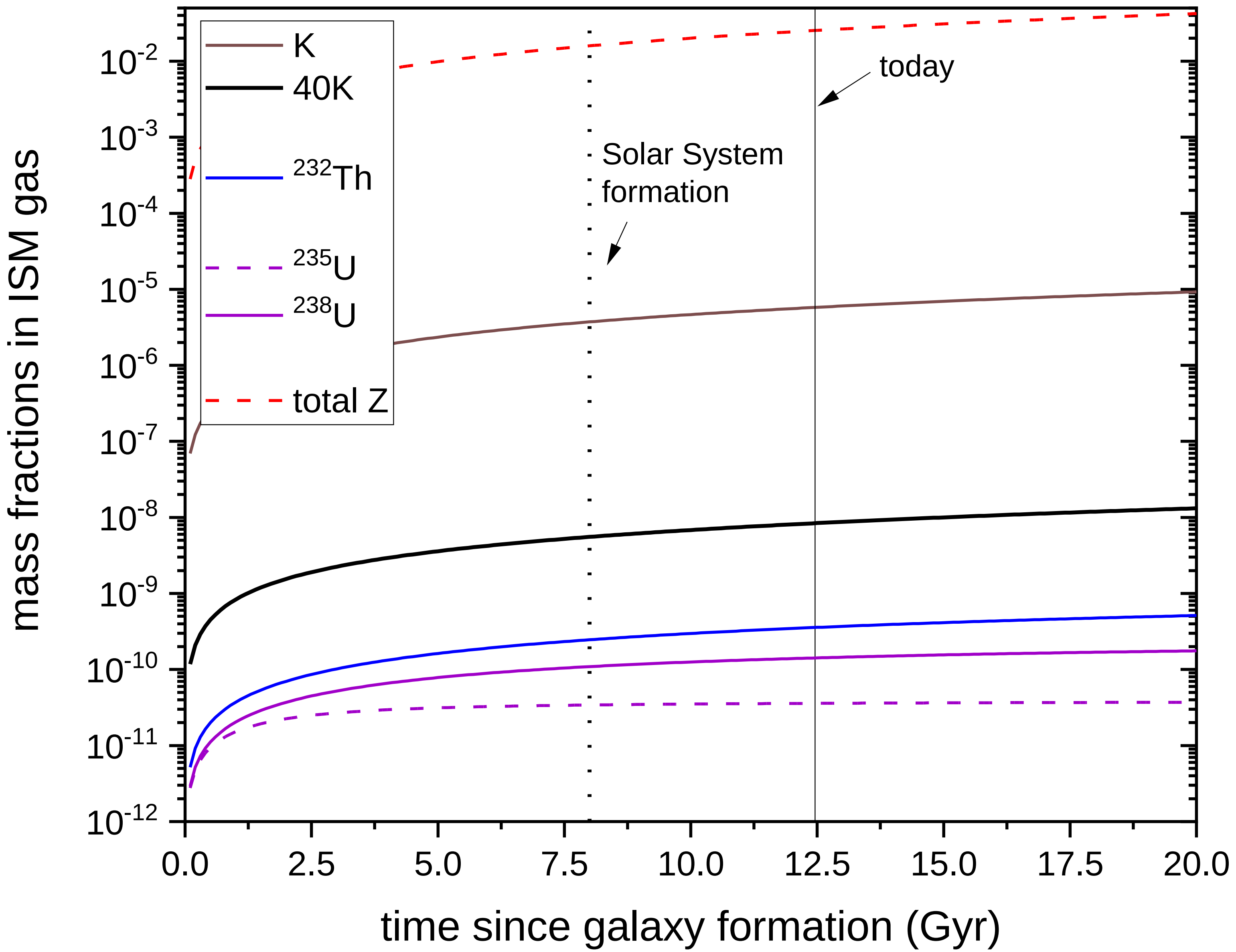

# Figure 19.

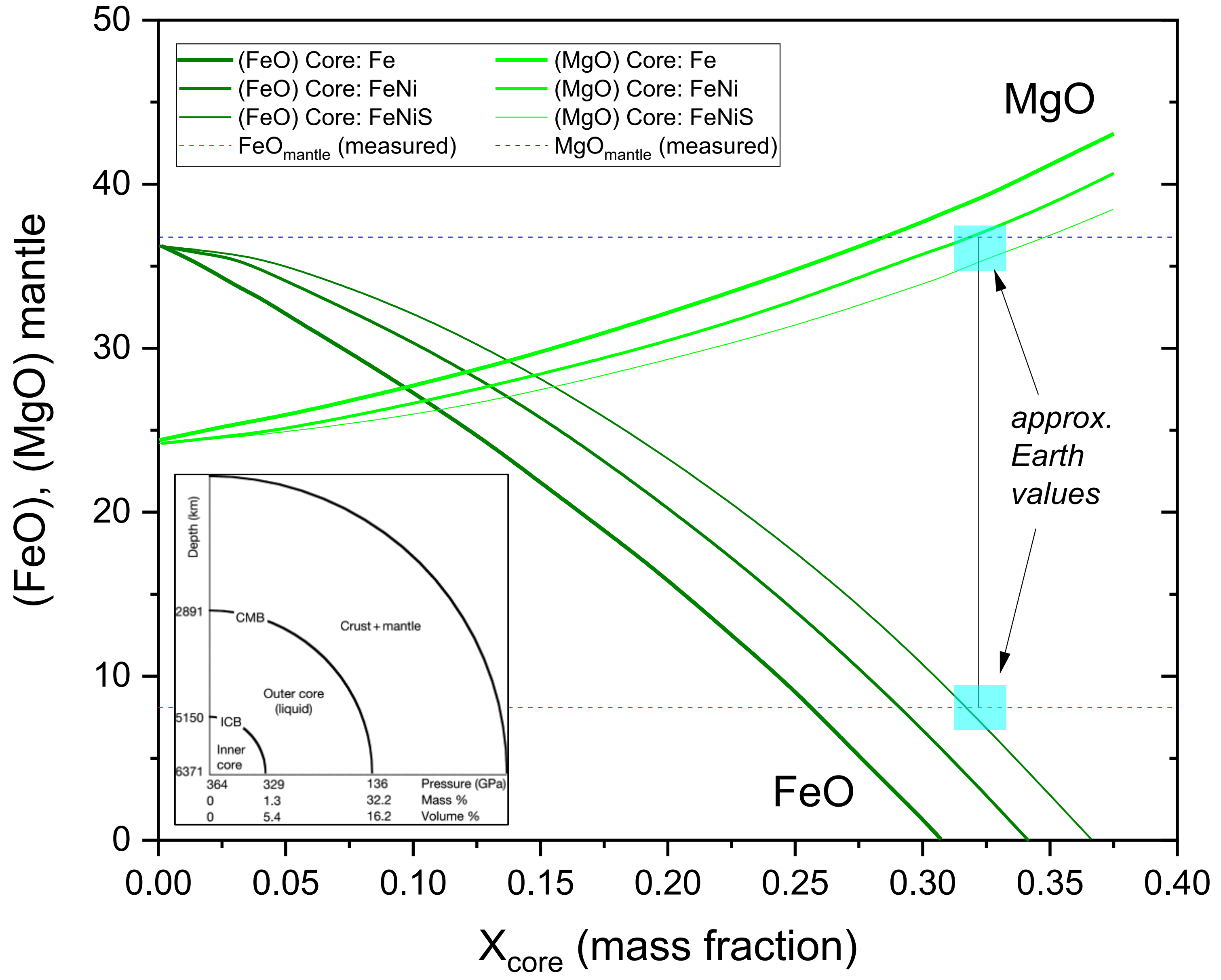

(FeO) Core: Fe
(FeO) Core: FeNi
(FeO) Core: FeNiS
FeO mantle (measured)
(MgO) Core: Fe
(MgO) Core: FeNi
(MgO) Core: FeNiS
MgO mantle (measured)
MgO
FeO
approx. Earth values
Depth (km)
2891
5150
6371
CMB
ICB
Crust + mantle
Outer core (liquid)
Inner core
364 329 136 Pressure (GPa)
0 1.3 32.2 Mass %
0 5.4 16.2 Volume %
(FeO), (MgO) mantle
X core (mass fraction)
0 10 20 30 40 50
0.00 0.05 0.10 0.15 0.20 0.25 0.30 0.35 0.40

**Figure 20.**

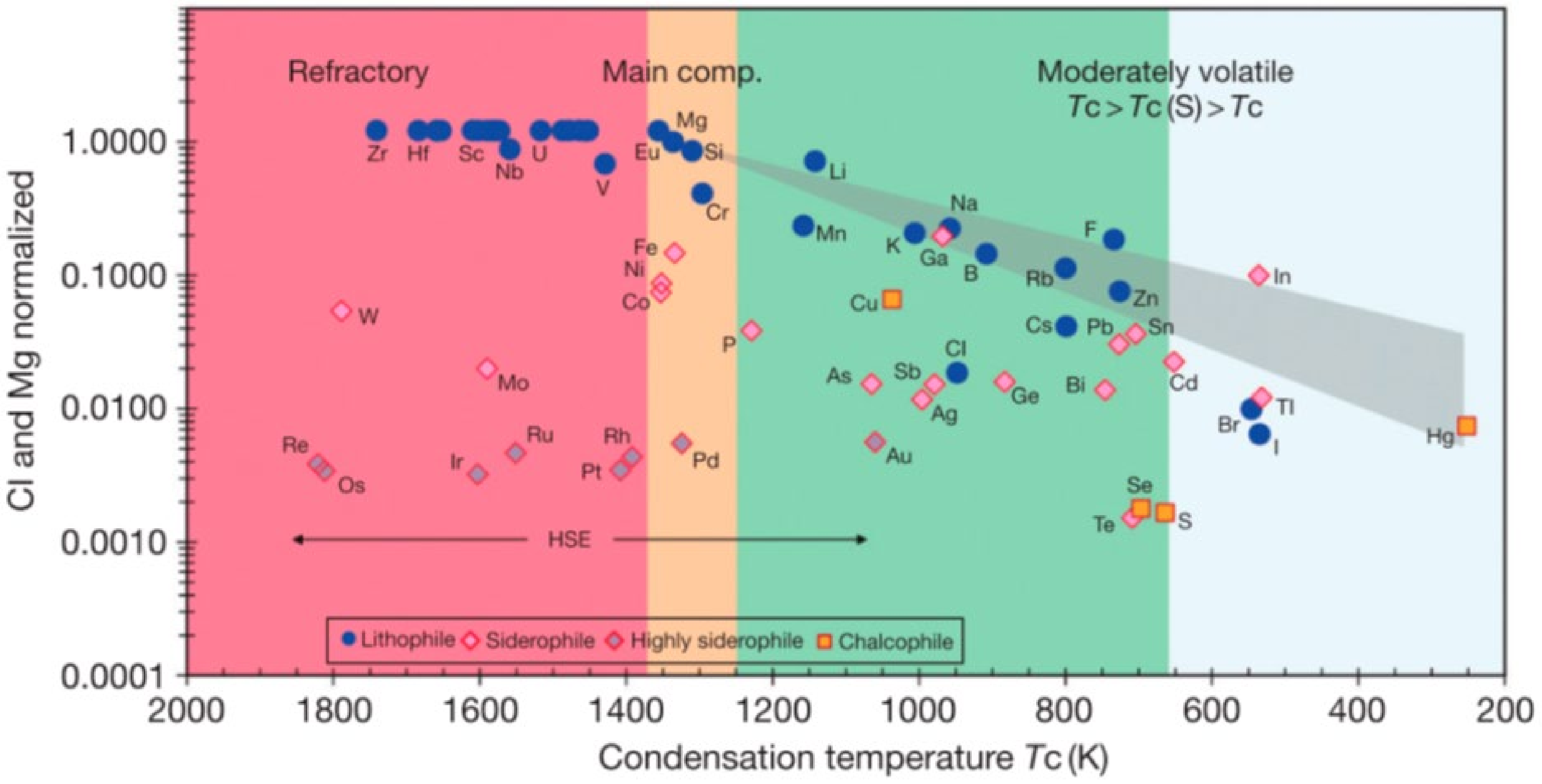

**Figure 21.**

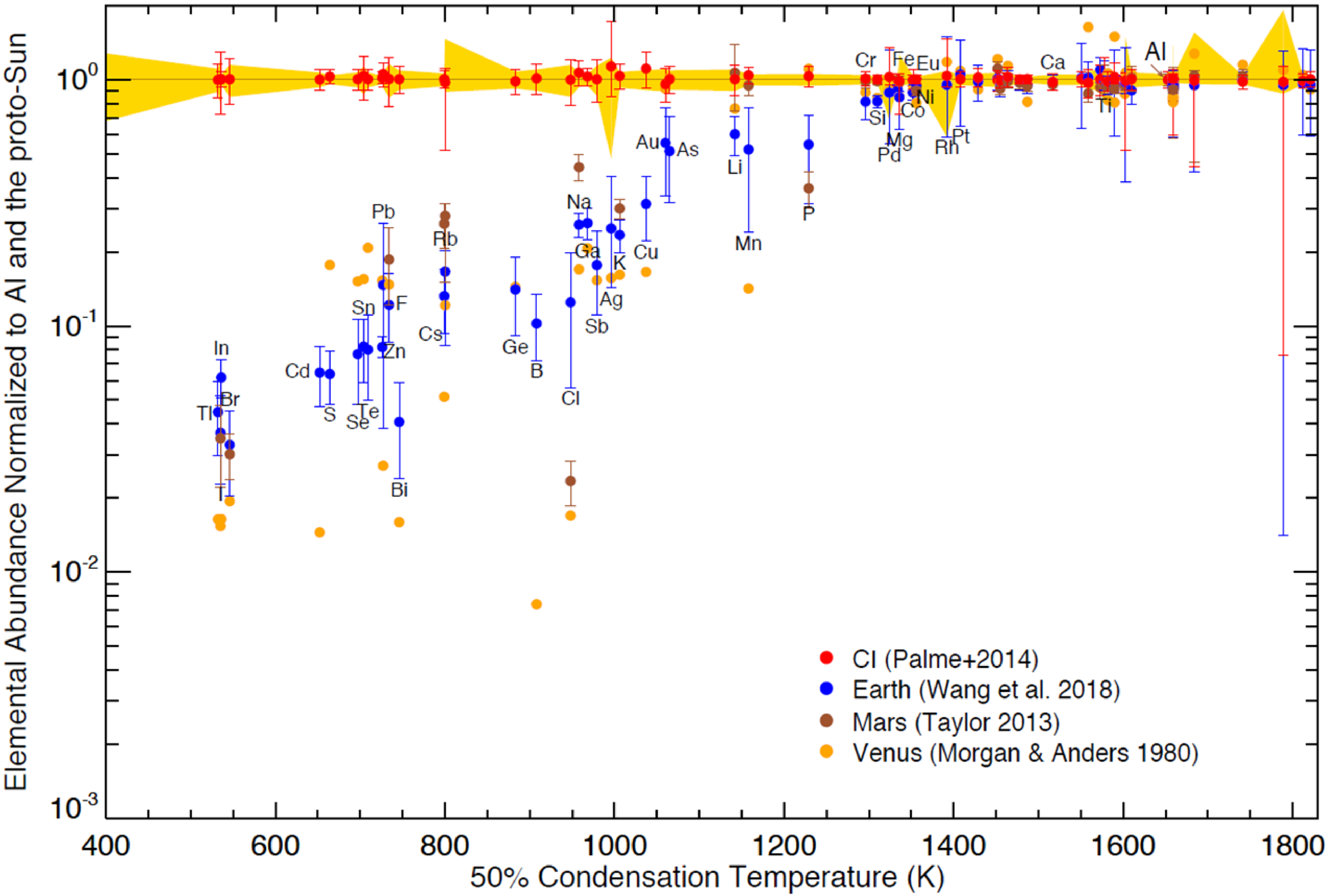

Elemental Abundance Normalized to Al and the proto-Sun
50% Condensation Temperature (K)
CI (Palme+2014)
Earth (Wang et al. 2018)
Mars (Taylor 2013)
Venus (Morgan & Anders 1980)
In
Tl
Br
I
Cd
S
Se
Te
Sn
Pb
Zn
F
Bi
Rb
Cs
Ge
B
Cl
Na
Ga
K
Ag
Sb
Cu
Au
As
Li
Mn
P
Cr
Si
Fe
Eu
Ni
Co
Mg
Pd
Rh
Pt
Ca
Ti
Al

**Figure 22.**

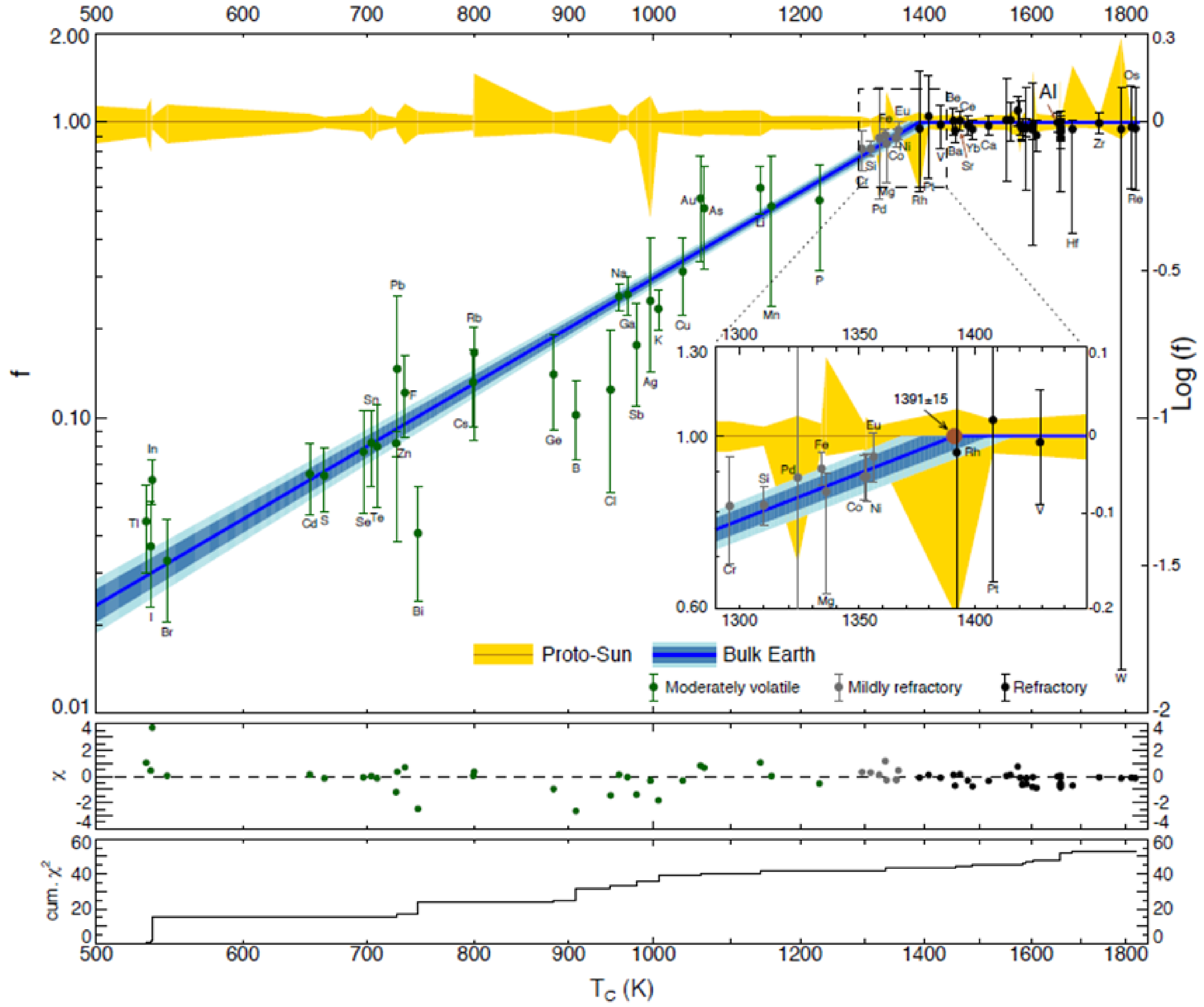